%% file: EADF_in_massive_MIMO.tex
\documentclass[10pt,a4paper,twocolumn]{IEEEtran} 
\normalsize
\usepackage[colorlinks,
linkcolor=black,
anchorcolor=black,
citecolor=black,
urlcolor=black,
]{hyperref}
\usepackage[hyphenbreaks]{breakurl}
\usepackage{diagbox}
\usepackage[english]{babel}
\usepackage{latexsym}
\usepackage{varioref}
\usepackage{mathrsfs}
\usepackage{amssymb}
\usepackage{array}
\usepackage{hhline}
\usepackage{dcolumn}
\usepackage{tabularx}
\usepackage{algorithm}
\usepackage{algorithmic}
\usepackage{calc}
\usepackage[usenames]{color}
\usepackage{amsmath,amssymb,bigdelim}
\usepackage{amsfonts}
\usepackage{t1enc}
\usepackage{graphicx}
\usepackage{pstool}
\usepackage{boxedminipage}
\usepackage{indentfirst}
\usepackage{booktabs}
\usepackage[noadjust]{cite}
\usepackage{caption}
\usepackage{multirow}
\usepackage{subfigure}
\usepackage[usenames]{color}
\usepackage{clrscode}
\usepackage{url}
\usepackage{setspace}
\usepackage{xcolor}
\usepackage{fancybox}

\makeatletter
\def\url@leostyle{%
  \@ifundefined{selectfont}{\def\UrlFont{\sf}}{\def\UrlFont{\small\ttfamily}}}
\makeatother
\input{newcommands_thesis2.tex}
\newcommand{\addnew}[1]{{\color{black} #1}}
\newcommand{\addnewfr}[1]{{\color{black} #1}}
\renewcommand{\baselinestretch}{1}

\providecommand{\texlivekeywords}[1]{\textbf{\textit{Index terms---}}#1}
\begin{document}

	\pagestyle{plain}
  \title{Enhanced \addnew{Effective Aperture Distribution Function for Characterizing Large-Scale Antenna Arrays}}
	\author{Xuesong Cai,~\IEEEmembership{Senior Member,~IEEE}, Meifang Zhu, Aleksei Fedorov, and Fredrik Tufvesson,~\IEEEmembership{Fellow,~IEEE}
		\thanks{
      This work has been funded by the Horizon Europe Framework Programme under the Marie Skłodowska-Curie grant agreement No.,101059091, the Swedish Research Council (Grant No. 2022-04691), the strategic research area ELLIIT, Excellence Center at Linköping — Lund in Information Technology, and Ericsson. (\textit{Corresponding author: X. Cai})  
    

      X. Cai, A. Fedorov and F. Tufvesson are with the Department of Electrical and Information Technology, Lund University, Lund, Sweden (email: xuesong.cai@eit.lth.se, aleksei.fedorov@eit.lth.se, fredrik.tufvesson@eit.lth.se). 

    M. Zhu is with Ericsson AB, Lund, Sweden (email: meifang.zhu@ericsson.com).
    
    }
}

\markboth{IEEE Transactions on Communications}%
{Submitted paper}

\maketitle \thispagestyle{plain}
\begin{abstract}

\addnewfr{Accurate characterization of large-scale antenna arrays is growing in importance and complexity for the fifth-generation (5G) and beyond systems, as they feature more antenna elements and require increased overall performance. The full 3D patterns of all antenna elements in the array need to be characterized because they are in general different due to construction inaccuracy, coupling, antenna array’s asymmetry, etc.
The effective aperture distribution function (EADF) can provide an analytic description of an antenna array based on a full-sphere measurement of the array in an anechoic chamber. However, as the array aperture increases, denser spatial samples are needed for EADF due to large distance offsets of array elements from the reference point in the anechoic chamber, leading to a prohibitive measurement time and increased complexity of EADF.   
In this paper, we present the EADF applied to large-scale arrays and highlight issues caused by the large array aperture. To overcome the issues, an enhanced EADF is proposed with a low complexity that is intrinsically determined by the characteristic of each array element rather than the array aperture. The enhanced EADF is validated using experimental measurements conducted at 27-30\,GHz frequency band with a relatively large planar array.}




\end{abstract}
\texlivekeywords{Antenna measurement, large-scale array, array characterization, and channel parameter estimation.}
\IEEEpeerreviewmaketitle









\section{Introduction}




Massive multiple-input multiple-output (MIMO) has been a key technology for fifth-generation (5G) and beyond communication systems operating at the lower-frequency, the millimeter-wave (mmWave) and/or the sub-THz bands \cite{THzmagazine,6736761,6824752,9390169}. Utilizing a large number of antennas, massive MIMO can facilitate applications such as multi-user beamforming \cite{6979962}, localization \cite{8741206}, sensing \cite{9482537}, etc. To enable those advanced applications, an accurate characterization of the large-scale antenna array is essential. \addnewfr{This means that the full 3D response pattern of every antenna element in the array is to be known to determine proper weights for beamforming.} Another typical case is to utilize large-scale antenna arrays for measurement-based spatial channel sounding, where the array effect is required to be decoupled from the pure propagation channels for accurate modeling of channels. 
\addnewfr{There exist different array schemes, including real antenna arrays, switched arrays and beam-scanning arrays \cite{THzmagazine}, to sound angular characteristics of channels. For real antenna arrays, the received signals of all antennas are measured simultaneously. For switched antenna arrays, only one antenna is activated at one time to achieve lower hardware complexity \cite{mmWaveSounder}. While for beam-scanning arrays, analog beamforming is usually exploited to scan one direction at one time, which is somehow equivalent to the switched antenna array by considering each beam as an ``virtual antenna element''. For all the cases, the radiation pattern of each antenna element (either real or virtual) must be accurately known, so that they can be decoupled to retrieve the pure propagation channel characteristics. It is worth noting that even though in some cases all the antenna elements are designed to be the same, it is still practically necessary to measure and characterize all of them due to such as coupling, construction inaccuracy and frame asymmetry.}

To characterize an antenna array, the array is in general placed in an anechoic chamber, where the \addnew{far-field} responses of the antenna elements are measured at discrete angle steps covering a whole 3D sphere, and more than one probe may be needed for a polarimetric characterization of the array. The measurement results are usually stored in a form of matrices for recovering the antenna responses in an arbitrary direction.  

Different algorithms have been proposed in the literature for reconstruction purposes, which can be classified into three main types. 
\textit{i)} The first type is to directly perform, either linear or non-linear, interpolations based on the measured data to recover the responses of unmeasured directions. Although straightforward, the main drawback of such algorithms is that they are nonanalytic. This means that algebraic analysis, e.g., gradient-based optimization, cannot be done. Moreover, these algorithms are not designed for compactly storing the measurements, meaning that the full-size matrices are always carried and used throughout the following signal processing steps. 
\textit{ii)} The second type of algorithms is based on the so-called spherical harmonics (SHs) \cite{Giovanni1,6175298,5345757,gia2019favest}. The SHs are orthogonal basis functions that are defined on a sphere. The coefficients of different SHs contained in the measurement data can be obtained by projecting the measurement data onto the SHs, i.e., a forward transform. Then the array responses can be recovered by an inverse transform according to the obtained coefficients, i.e., a linear combination of the SHs. It is worth noting that if the measurement data is sparse in the transformed domain, one can only store the principal components to compress the measurement data. 
\textit{iii)} The third type is the so-called effective aperture distribution function (EADF) \cite{EADF,Landmann2012,9723279}. Essentially, 2D discrete Fourier transforms (DFTs) are applied to the measurement patterns to obtain spectra in the spatial-frequency domains that can be utilized afterwards to recover antenna patterns via inverse DFTs. This approach has several advantages. For example, similarly to using SHs, principal spatial-frequency components can be recognized to compress the measurement data. Moreover, both the forward and inverse DFTs can be performed efficiently in hardware. The EADF also can be extended for wideband characterization with 3D DFT by further considering the frequency-domain \cite{Full3d}.

Although EADF is a promising technique, there is an issue when applied to large-scale antenna arrays. With a large array aperture, in terms of wavelengths, the radiation patterns of the antenna elements measured in the anechoic chamber can have very high local spatial frequencies. 
For future THz communications with ultra massive MIMO \cite{9216613}, this issue will be of increased importance.  
Moreover, due to the deep integration of antennas in devices, a large aperture of the array can also be caused by the whole electromagnetic structure surrounding the antenna array. For example, we need to consider a car or car roof and an array installed on it as a new ``array'' whose aperture can be very large, and such consideration is necessary in practice since the car body can influence the array responses significantly \cite{9280910}. 
The high local spatial frequencies caused by the large array aperture require very dense angular samples on the sphere surface, resulting in prohibitive array measurement efforts in chambers and high computation complexities. Otherwise, erroneous reconstruction of the array responses can happen.

As a critical part of 5G and beyond communications, accurate and complexity-efficient EADF-based characterization of large-scale antenna arrays is essential. However, to the authors' best knowledge, the problem of the conventional EADF confronted in large-scale antenna arrays has not been addressed in the literature. Therefore, we aim at proposing a novel and low-complexity EADF for characterizing large-scale antenna arrays by solving the aforementioned issue. The main novelty and contributions of the paper include:
\begin{enumerate}
    \item The EADF for characterizing antenna arrays is discussed, and the problem of its application to antenna arrays with large apertures is analyzed theoretically. 
    \item An enhanced EADF is proposed to overcome the problem. The main principle of the enhancement is to estimate the phase centers of the antenna elements so that the high spatial-frequency components can be sub-sampled. In such a way, the angle steps for spatial sampling are almost only related to the intrinsic patterns of the elements (rather than the aperture), which can significantly reduce the time consumption of the 
    antenna array measurement campaigns and the sizes of radiation pattern
    matrices, i.e., the complexity of further signal processing. 
    \item A mmWave antenna array is measured in an anechoic chamber at 27-30\,GHz. The enhanced EADF is evaluated, and the proposed method is verified using the measurement data. 
\end{enumerate}

The rest of the paper is structured as follows. 
The EADF for antenna arrays is discussed in Sect.\,\ref{sect:eadf_for_array}. 
Sect.\,\ref{sect:eadf_issue} analyzes the issue of the conventional EADF when applied to large-scale antenna arrays.
 In Sect.\,\ref{sect:enhanced_eadf}, the enhanced EADF is proposed, and the measurement campaign is introduced in Sect.\,\ref{sec:measurement_verification}. 
 Finally, conclusive remarks are given in Sect.\,\ref{sect:conclusions}. 
Throughout this paper, we use italic letters to denote scalars, bold letters in lower case for vectors, and bold letters in upper case for matrices. The function $\exp()$ denotes the element-wise exponential function, and $j=\sqrt{-1}$. In addition, $\{\cdot\}^{T}$, $\text{vec}(\cdot)$, $\otimes$ and $\odot$ 
indicate the matrix transposition, matrix vectorization, Kronecker product and Hadamard product (element-wise product), respectively. \addnewfr{In addition, the inner product of two vectors $\mathbf a$ and $\mathbf b$, i.e., $\mathbf a \cdot \mathbf b$, is equivalently written as $\mathbf a^T \mathbf b$.} 

\section{The EADF for an Antenna Array\label{sect:eadf_for_array}}

\begin{figure}
  \centering
  \psfrag{Anechoic Chamber}[l][l][0.8]{Inside an anechoic chamber}
  \psfrag{X}[c][c][0.6]{X}
  \psfrag{Y}[c][c][0.6]{Y}
  \psfrag{Z}[c][c][0.6]{Z}
  \psfrag{O}[c][c][0.6]{O}
  \psfrag{t}[c][c][0.7]{$\theta$}
  \psfrag{phi}[l][l][0.7]{$\phi$}
  \psfrag{R}[c][c][0.6]{$r$}
  \psfrag{Probe antenna}[c][c][0.6]{Probe antenna}
  \psfrag{pth antenna element}[l][l][0.6]{The $p$-th element with its phase}
  \psfrag{with}[l][l][0.6]{center at ($ x^{(p)}$, $ y^{(p)}$, $z^{(p)}$)}
  \includegraphics[width=0.45\textwidth]{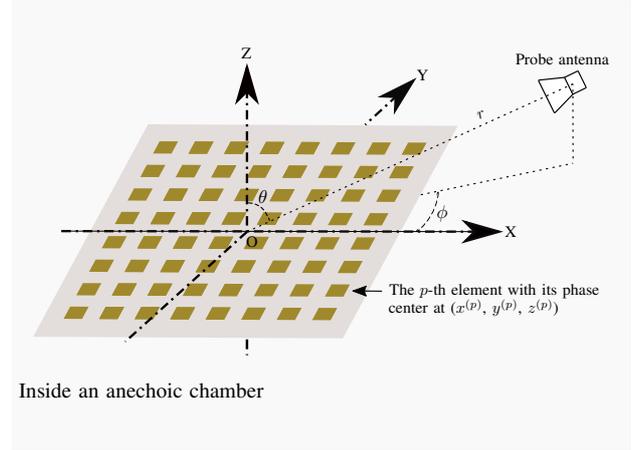}{}
  \caption{A schematic of (large-scale) antenna array characterization in an anechoic chamber. \label{fig:array_measurement}}
 \end{figure}

Fig.\,\ref{fig:array_measurement} illustrates a schematic of antenna array characterization in an anechoic chamber. The antenna array consisting of $P$ elements is placed around the center ``O'' of a sphere with radius $r$. On the sphere surface, a calibrated probe antenna is used to measure the responses/patterns of all the $P$ antenna elements at different directions $(\theta, \phi)$s, where $\theta \in [0, \pi]$ is the zenith angle with $\theta=0$ corresponding to the positive $z$-axis, and $\phi \in [0, 2\pi]$ is the azimuth angle with $\azimuth=0$ indicating the positive $x$-axis. The radius $r$ is set much larger than the Fraunhofer distance \cite{9104014,8713575} of the antenna array to guarantee a far-field characterization. In practice, there are different ways to equivalently realize the measurements as illustrated in Fig.\,\ref{fig:array_measurement}. For example, one can rotate either or both the array under test or/and the probe antenna. A reflector can also be used to create far-field conditions in an anechoic chamber with limited dimensions \cite{1373997}.

\begin{figure}
    \centering
    \psfrag{zenith}[c][c][0.6]{$\theta$ $[^\circ]$}
    \psfrag{azimuth}[c][c][0.6]{$\phi$ $[^\circ]$ }
    \psfrag{Gain [dB]}[c][c][0.6]{Gain [dB]}
    \subfigure[]{\includegraphics[width=0.49\textwidth]{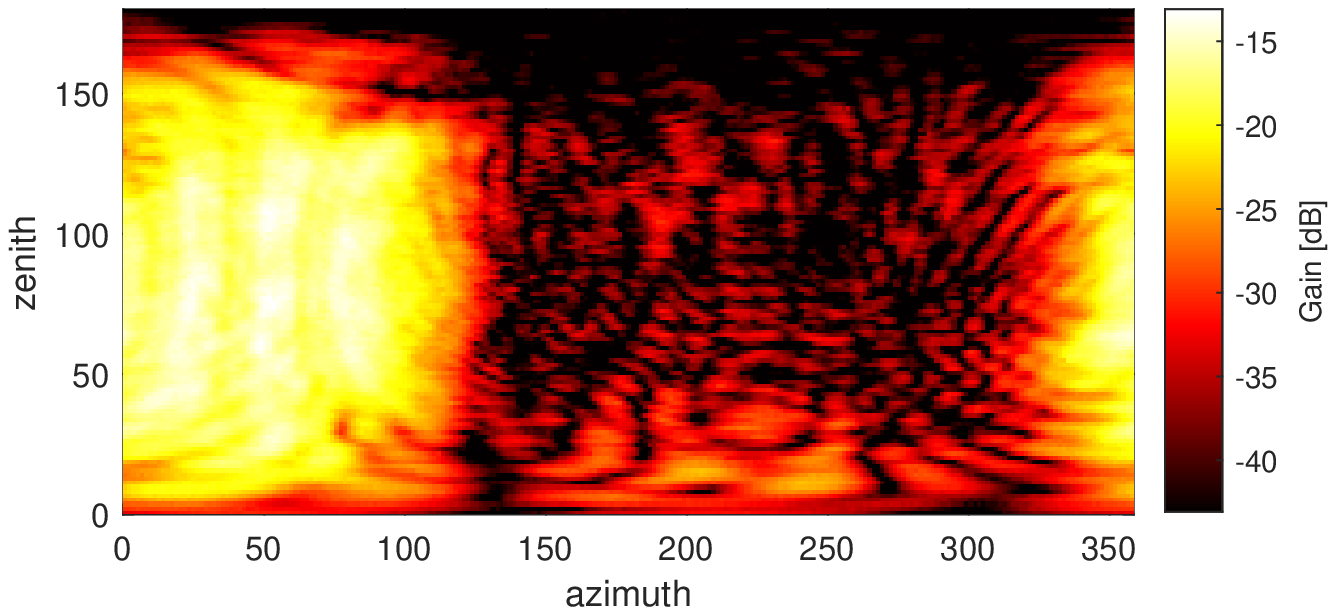}\label{fig:pattern_and_eadf_a}}
    \psfrag{zenith}[c][c][0.6]{$\theta$ $[^\circ]$}
    \psfrag{azimuth}[c][c][0.6]{$\phi$ $[^\circ]$ }
    \psfrag{Gain [dB]}[c][c][0.6]{Gain [dB]}
    \subfigure[]{\includegraphics[width=0.24\textwidth]{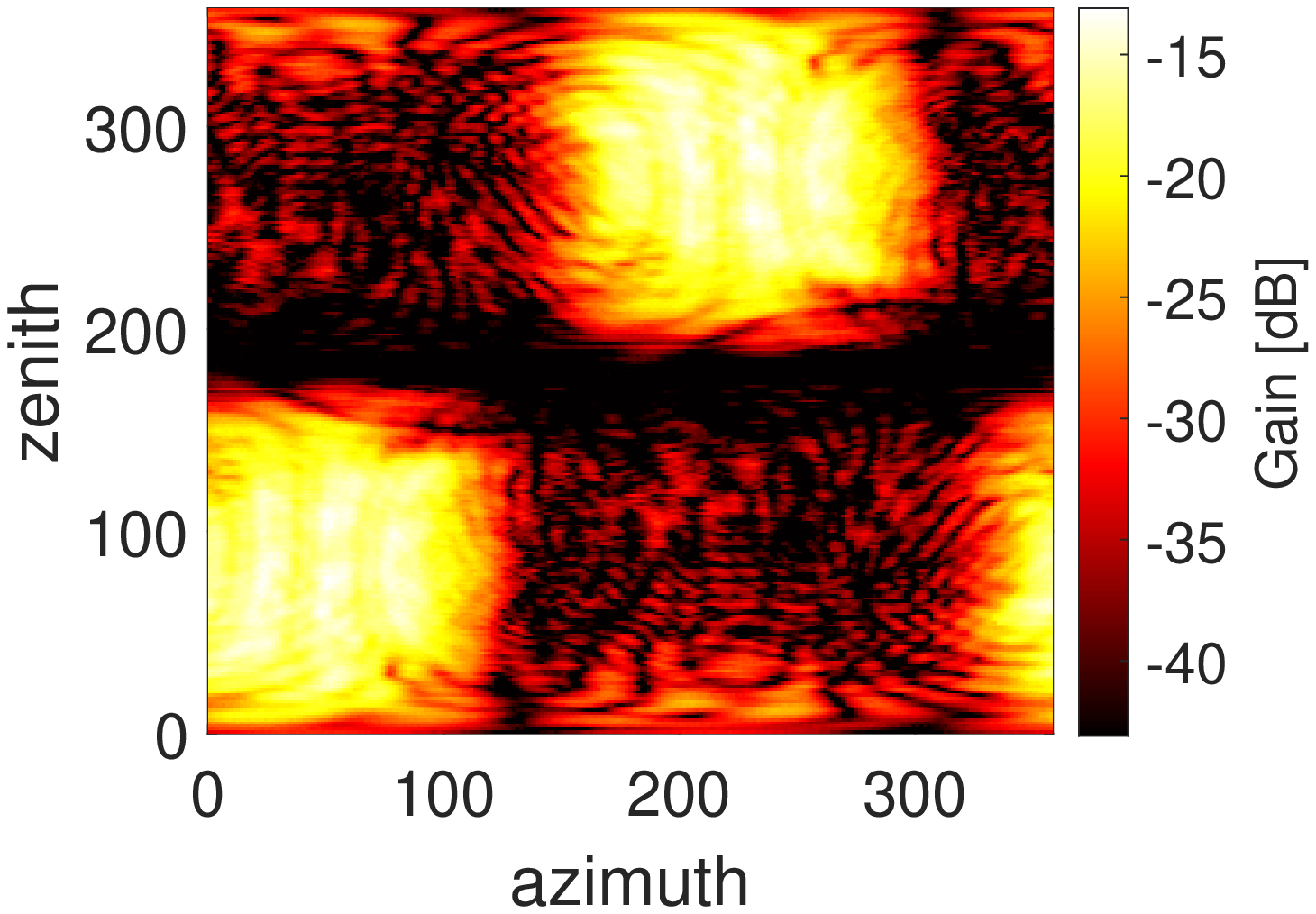}\label{fig:pattern_and_eadf_b}}
    \psfrag{zenith frequency}[c][c][0.6]{Zenithal spatial frequency}
    \psfrag{azimuth frequency}[c][c][0.6]{Azimuthal spatial frequency}
    \psfrag{Power [dB]}[c][c][0.6]{Normalized power [dB]}
    \subfigure[]{\includegraphics[width=0.24\textwidth]{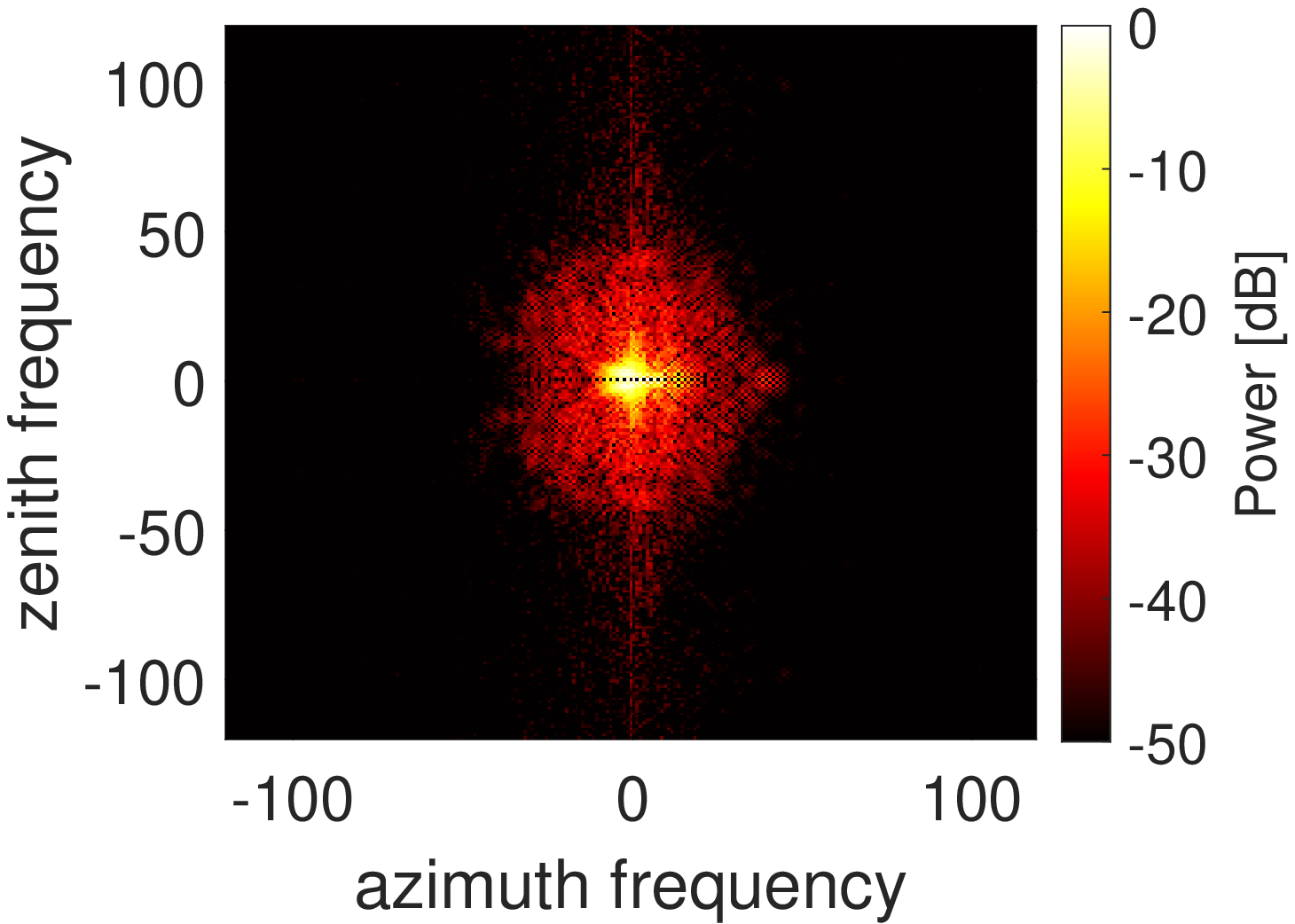}\label{fig:pattern_and_eadf_c}}
    \caption{The measured 3D pattern and the corresponding EADF of a patch antenna at 28.5\,GHz. (a) The measured 3D pattern $\bold A$. (b) Extended pattern $\bold C$. (c) The obtained EADF $\bold Q$ with normalized power. \label{fig:pattern_and_eadf}}
   \end{figure}

The array responses are generally measured at discrete angular support. Specifically, the measured 3D radiation pattern of the $p$-th element can be denoted as $\mathbf A^{(p)}\in \mathbb{C}^{(M+1)\times 2N}$. The rows and columns of $\bold A^{(p)}$ correspond to $\boldsymbol{\theta} = [0, \frac{\pi}{M}, \cdots, \pi]^{T}$ and $\boldsymbol{\phi} = [0, \frac{\pi}{N}, \cdots, \frac{(2N-1)\pi}{N}]^{T}$, respectively. 
As an example, Fig.\,\ref{fig:pattern_and_eadf_a} illustrates a measured $\mathbf A$ at 28.5\,GHz for a patch antenna element of an array with $M+1=121$ and $2N=240$, respectively. 
That is, the angle steps are 1.5 degrees for both $\boldsymbol{\theta}$ and $\boldsymbol{\phi}$. It can be observed that the patch element can effectively illuminate a certain area. The EADF for the measured pattern is essentially the 2D-DFT of $\bold A^{(p)}$. 
\addnew{However, directly applying 2D-DFT to $\bold A^{(p)}$ will lead to distortions of the recovered patterns in the local regions close to the north and south poles.\footnote{\addnew{Considering one azimuth cut (a half circle) of the measured pattern from the north to the south pole, its discrete 1D-DFT spectrum corresponds to a periodic ``time-domain'' (spatial-domain) function. This means that the recovered continuous pattern is almost ``... [north-pole to south-pole] [north-pole to south-pole] [north-pole to south-pole] ...''. Looking at one period, its neighboring value to the left is from the south pole, and to the right is from the north pole. The discontinuities will lead to the recovered pattern close to the south/north pole being distorted. To avoid this issue, the whole circle at every azimuth cut should be exploited to obtain the EADF.}}}
Therefore, the original $\bold A^{(p)}$ needs to be extended to $\bold C^{(p)} \in \mathbb{C}^{2M\times 2N}$, where the rows and columns of $\bold C^{(p)}$ correspond to $\boldsymbol{\theta} = [0, \frac{\pi}{M}, \cdots, \frac{(2M-1)\pi}{M}]^{T}$ and $\boldsymbol{\phi} = [0, \frac{\pi}{N}, \cdots, \frac{(2N-1)\pi}{N}]^{T}$, respectively. The new $\bold C^{(p)}$ can be constructed by concatenating $\bold A^{(p)}$ and $\bold B^{(p)}$, where $\bold B^{(p)} \in \mathbb{C}^{(M-1)\times 2N}$ is obtained by circularly shifting the columns of $\bold A^{(p)}$ by $N$ steps (an angle of $\pi$), removing the first and last rows,  and finally reversing the order of all rows. The procedure can be formulated as  
\begin{equation}
  \begin{aligned}
   \bold C^{(p)} =  \begin{bmatrix}
      \bold A^{(p)} \\
      \bold B^{(p)}
      \end{bmatrix}
   \end{aligned}
   \label{eq:extending_3d_pattern}
   \end{equation} 
   with 
   \begin{equation}
    \begin{aligned}
     \bold B^{(p)} = \begin{bmatrix}
      \bold e_{M}^{T} \\
       \vdots \\
        \bold e_2^{T}\\
      \end{bmatrix} \bold A^{(p)} \begin{bmatrix}
        \addnew{\bold 0_N} & \bold I_N  \\
        \bold I_N & \addnew{\bold 0_N}
        \end{bmatrix},
     \end{aligned}
     \label{eq:aprime} 
     \end{equation} where $\bold e_m \in \mathbb{C}^{(M+1)\times 1}$ is a column vector with only its $m$-th entry as one and all the other entries as zero, $\bold I_N$ is an identity matrix of order $N$, and \addnew{$\bold 0_N$ is an $N\times N$ zero matrix}. As an example, Fig.\,\ref{fig:pattern_and_eadf_b} illustrates the constructed $\bold C$ for the patch element. The EADF $\bold Q^{(p)} \in \mathbb{C}^{ 2M\times 2N}$ can then be calculated according to $\bold C^{(p)}$. Specifically, the $(k,\ell)$-th entry $q_{k\ell}^{(p)}$ of $\bold Q^{(p)}$ is formulated as 
\begin{equation}
    \begin{aligned}
      q_{k\ell}^{(p)} = \boldsymbol{\mu}_k^{T} \bold C^{(p)} \boldsymbol{\mu}_\ell,
     \end{aligned}
     \label{eq:eadf_calculation}
     \end{equation} where 
     \begin{equation}
      \begin{aligned}
        \boldsymbol{\mu}_k = \exp(-j\pi\frac{(k-1-M)}{M} [0,\cdots,2M-1]^{T})
       \end{aligned}
       \end{equation} and 
       \begin{equation}
        \begin{aligned}
          \boldsymbol{\mu}_\ell = \exp(-j\pi\frac{(\ell-1-N)}{N} [0,\cdots,2N-1]^{T}).
         \end{aligned}
         \end{equation} In such a way, the zero spatial frequency component is located in the middle of $\bold Q^{(p)}$. Fig.\,\ref{fig:pattern_and_eadf_c} illustrates the normalized EADF $\bold Q$ for the patch element as illustrated in Fig.\,\ref{fig:pattern_and_eadf_a}. It can be observed that the main power of $\bold Q$ is concentrated in the lower spatial frequency region. This is an essential prerequisite so that the EADF can be exploited to recover/interpolate the radiation pattern (response) of an antenna at an arbitrary direction $(\theta, \phi)$. Otherwise, aliasing caused by high spatial frequency components can probably lead to failures of the interpolation of radiation patterns if $\bold A^{(p)}$s were not sampled with adequate angular resolution.  

Utilizing the EADF, the response $a^{(p)}(\theta, \phi)$ at an arbitrary direction $(\theta, \phi)$ for the $p$-th element can be recovered as
\begin{equation}
  \begin{aligned}
    \hat{a}^{(p)}(\theta, \phi) & = \boldsymbol{\omega}_\theta^{T} \bold Q^{(p)} \boldsymbol{\omega}_\phi\\
                         & = \text{vec}^{T}(\bold Q^{(p)})  \boldsymbol{\omega}_{\theta\phi} 
   \end{aligned}
   \end{equation}
where $\boldsymbol{\omega}_{\theta\phi} = {\boldsymbol{\omega}_\theta \otimes \boldsymbol{\omega}_\phi}$ with 
\begin{equation}
  \begin{aligned}
    \boldsymbol{\omega}_\theta = \exp(j\theta [-M,\cdots,M-1]^{T})
   \end{aligned}
   \end{equation} and 
   \begin{equation}
    \begin{aligned}
      \boldsymbol{\omega}_\phi = \exp(j\phi [-N,\cdots,N-1]^{T}).
     \end{aligned}
     \end{equation} 
Therefore, the array response $\bold g(\theta, \phi)$ with its $p$-th entry as $\hat{a}^{(p)}(\theta, \phi)$ can be formulated as 
     \begin{equation}
      \begin{aligned}
        \bold g(\theta, \phi) = \begin{bmatrix}
          \text{vec}^{T}(\bold Q^{(1)})  \\
           \vdots \\
           \text{vec}^{T}(\bold Q^{(P)}) \\
          \end{bmatrix} \boldsymbol{\omega}_{\theta\phi}. 
       \end{aligned}
       \label{eq:singlepol}
       \end{equation}    
Furthermore, considering both horizontally (H-) and vertically (V-) polarized responses of the array, \eqref{eq:singlepol} can be extended to
\begin{equation}
  \begin{aligned}
    \bold G(\theta, \phi) & = \begin{bmatrix} 
      \bold g_{\text{H}}(\theta, \phi) & \bold g_{\text{V}}(\theta, \phi) 
      \end{bmatrix} \\
   & = \begin{bmatrix} 
      \text{vec}^{T}(\bold Q_{\text{H}}^{(1)})  &  \text{vec}^{T}(\bold Q_{\text{V}}^{(1)}) \\
       \vdots &  \vdots \\
       \text{vec}^{T}(\bold Q_{\text{H}}^{(P)}) &  \text{vec}^{T}(\bold Q_{\text{V}}^{(P)})  \\
      \end{bmatrix} \begin{bmatrix} 
        \boldsymbol{\omega}_{\theta\phi}  & \mathbf 0 \\
        \mathbf 0 &  \boldsymbol{\omega}_{\theta\phi} \\
        \end{bmatrix}, 
   \end{aligned}
   \label{eq:poloreadf}
   \end{equation} which is a complete polarimetric characterization of the array by exploiting the EADF. Note that H-polarized and V-polarized probes are required in the array measurements as illustrated in Fig.\,\ref{fig:array_measurement} to obtain polarized $\bold Q_{\text{H}}^{(p)}$s and $\bold Q_{\text{V}}^{(p)}$s in \eqref{eq:poloreadf}. It is also worth noting that \eqref{eq:poloreadf} provides an analytic model of the array, i.e. an analytic communication channel model \cite{9664391}, and the advantage to enable gradient-based optimization, e.g., to maximize the likelihood functions of channel parameters of propagation paths to achieve high-resolution estimation results \cite{richter2005estimation}.

 \section{The issue of EADF for large-scale arrays\label{sect:eadf_issue}}


For the characterization of a single antenna in an anechoic chamber, it is possible to align the phase center of the antenna to the sphere center ``O'' as illustrated in Fig.\,\ref{fig:array_measurement}. However, for an antenna array with a certain number of antenna elements, it is inevitable that there are misalignments between the sphere center and the phase centers of different antenna elements. Moreover, a fixture is usually needed in practice to hold and/or rotate the antenna array in the chamber, which could lead to further misalignments.
As illustrated in Fig.\,1, let us consider the $p$-th antenna element whose phase center is located at $( x^{(p)},  y^{(p)},  z^{(p)} )$ in the Cartesian coordinate system. 
If omitting the additive noises\footnote{Note that there always exists noise in the measured data. We omit the noise term to make the following analysis and formulas more concise without losing the essence.}, the measured radiation pattern of the $p$-th antenna element can be formulated as   
\begin{equation} 
  \begin{aligned}
   a^{(p)}(\theta, \phi) = a_\text{0}^{(p)}(\theta, \phi) \exp(j 2\pi \lambda^{-1}{\mathbf u^T_{\theta, \phi} \mathbf d^{(p)}}   ),
  \end{aligned}
  \label{eq:measured_pattern}
  \end{equation} 
  where $\lambda$ is the wavelength, 
  $\mathbf u_{\theta, \phi}$ is a unit vector in the direction $(\theta, \phi)$ and written as  
  \begin{equation} 
    \begin{aligned}
      \mathbf u_{\theta, \phi} = [\sin\theta\cos \phi, \sin \theta \sin\phi, \cos\theta]^{T}, 
    \end{aligned}
    \end{equation} and $\mathbf d^{(p)}$ is the vector from the sphere center to the phase center of the $p$-th antenna element which is 
    \begin{equation} 
      \begin{aligned}
        \mathbf d^{(p)} = [ x^{(p)},  y^{(p)},  z^{(p)} ]^{T}, 
      \end{aligned}
      \end{equation} and $a_\text{0}^{(p)}(\theta, \phi)$ is the 
       response obtained if the phase-center of the antenna element is aligned to the sphere center.

To understand the issue of the phase-center misalignment, let us assume that the pattern of the $p$-th antenna element is omnidirectional, i.e., $a_\text{0}^{(p)}(\theta, \phi) = 1$. Thus we have 
\begin{equation}
\begin{aligned}
  a^{(p)}(\theta, \phi) =  \exp(j 2\pi \lambda^{-1} {\mathbf u^T_{\theta, \phi} \mathbf d^{(p)}}   ).
 \end{aligned}
 \label{eq:effectofoffset}
 \end{equation}
The first order approximation for $\zeta_{\theta, \phi}^{(p)} = \lambda^{-1} \mathbf u^T_{\theta, \phi} \mathbf d^{(p)}$ in \eqref{eq:effectofoffset} at $(\theta_0, \phi_0)$ is 
\begin{equation}
  \begin{aligned}
    \zeta_{\theta, \phi}^{(p)}  
   &  \approx \frac{\partial{\zeta_{\theta, \phi}^{(p)}}}{\partial{\theta}} \Biggr|_{\theta_0} (\theta - \theta_0) +  \frac{\partial{\zeta_{\theta, \phi}^{(p)}}}{\partial{\phi}} \Biggr|_{ \phi_0} (\phi - \phi_0)   + \zeta_{\theta_0, \phi_0}^{(p)}  \\
   & =    f_{\theta|\theta_0, \phi_0}^{(p)} \theta + f_{\phi|\theta_0, \phi_0}^{(p)} \phi   + \beta_{\theta_0, \phi_0}
   \end{aligned}
   \label{eq:approximation}
   \end{equation} where
     \begin{equation}
   \begin{aligned}
    &f_{\theta|\theta_0, \phi_0}^{(p)} = \frac{x^{(p)} \cos\theta_0 \cos\phi_0 + y^{(p)} \cos \theta_0 \sin\phi_0 - z^{(p)} \sin \theta_0}{\lambda}\\ 
       & f_{\phi|\theta_0, \phi_0}^{(p)} = \frac{- x^{(p)}\sin\theta_0 \sin\phi_0 + y^{(p)} \sin\theta_0 \cos\phi_0}{\lambda}, \\
     \end{aligned}
     \end{equation} 
     and $\beta_{\theta_0, \phi_0}$ collects all the residual terms that are only related to $\phi_0$ and $\theta_0$. 
   Therefore, $a^{(p)}(\phi,\theta)$ can be approximated as 
   \begin{equation}
    \begin{aligned}
      a^{(p)}(\phi,\theta) \approx  \exp(  j 2\pi (f_{\theta|\phi_0,\theta_0} \theta  + f_{\phi|\phi_0,\theta_0} \phi  + \beta_{\theta_0, \phi_0}) ).  
     \end{aligned}
     \label{eq:effectofoffset_approximated}
     \end{equation} It can be observed from \eqref{eq:effectofoffset_approximated} that in a spatial region near to $(\theta_0,\phi_0)$, the local ``spatial frequencies'' of the measured radiation pattern $a^{(p)}$ with respect to zenith and azimuth are approximately $f_{\theta|\phi_0,\theta_0}^{(p)}$ and $f_{\phi|\phi_0,\theta_0}^{(p)}$, respectively. Therefore, the maximum absolute  $f_{\phi|\phi_0,\theta_0}^{(p)}$ across the whole sphere can be derived as
     \begin{equation}
        \begin{aligned}
        & f_{\phi,\text{max}}^{(p)}  \\
        & = \max_{\phi_0,\theta_0}  ~  \lambda^{-1} \biggl|\sin(\phi_0 - \tan^{-1}\frac{y^{(p)}}{x^{(p)}})\biggr| \sqrt{ ({x^2}^{(p)} + {y^2}^{(p)}) \sin^2\theta_0} \\
        & = \lambda^{-1} \sqrt{ {x^2}^{(p)} + {y^2}^{(p)}}.
      \end{aligned}
    \end{equation} 
    Similarly, it can be known that 
       \begin{equation}
        \begin{aligned}
        f_{\theta,\text{max}}^{(p)} & = \lambda^{-1} \sqrt{ {x^2}^{(p)} + {y^2}^{(p)} +   {z^2}^{(p)}}. 
      \end{aligned}
      \label{eq:max_zenith_step}
    \end{equation} According to the Nyquist sampling theorem, the azimuth and zenith steps for the $p$-th antenna element during the measurements in the anechoic chamber should be no larger than $\frac{1}{2f_{\phi,\text{max}}^{(p)}} $ and $\frac{1}{2f_{\theta,\text{max}}^{(p)}}$ radians to avoid aliasing in the EADF, so that a correct reconstruction of $a^{(p)}$ at arbitrary directions can be achieved using EADF according to \eqref{eq:poloreadf}.  For a large-scale antenna array, the maximum admissible angle steps are determined by the maximum ${2f_{\phi,\text{max}}^{(p)}} $ and ${2f_{\theta,\text{max}}^{(p)}}$ of all the $P$ elements. Due to the large size, in terms of $\lambda$, of the antenna array, the admissible angle steps could be quite small, which results in a time-consuming and expensive array measurement in the anechoic chamber, and a computationally complex EADF characterization of the antenna array. It is also worth noting that for realistic antenna elements with a non-omnidirectional $a_\text{0}^{(p)}$, the requirement for angle steps can be even more demanding.

\section{The proposed enhanced EADF\label{sect:enhanced_eadf}}

In this section, we propose an enhanced EADF to characterize large-scale arrays, and the required angle steps are intrinsically limited by the characteristic of $a_\text{0}^{(p)}$ and irrespective of the array sizes. 

\subsection{The principle of the enhanced EADF}
Let us consider that the phase centers of the $P$ antenna elements are known or can be estimated as located at $( \hat{x}^{(p)},  \hat{y}^{(p)},  \hat{z}^{(p)} )$. That is, the vector connecting the sphere center and the estimated phase center of the $p$-th antenna element is 
$ \hat{\mathbf d}^{(p)}$ with 
\begin{equation} 
  \begin{aligned}
     \hat{\mathbf d}^{(p)} = [ \hat{x}^{(p)},  \hat{y}^{(p)},  \hat{z}^{(p)} ]^{T}. 
  \end{aligned}
  \end{equation} To enhance the EADF, the measured ${a}^{(p)}(\theta, \phi)$ is first processed as 
\begin{equation}
  \begin{aligned}
    \tilde{a}^{(p)}(\theta, \phi)  & =  a^{(p)}(\theta, \phi) \exp(-j 2\pi \lambda^{-1} \mathbf u^T_{\theta, \phi} \hat{\mathbf d}^{(p)}    ) \\
    &  = a_\text{0}^{(p)}(\theta, \phi) \exp(j 2\pi \lambda^{-1} \mathbf u^T_{\theta, \phi} \tilde{\mathbf d}^{(p)}    ).
   \end{aligned}
   \label{eq:compensation}
   \end{equation} where  
   \begin{equation}
    \begin{aligned}
      \tilde{\mathbf d}^{(p)}  & =   \mathbf d^{(p)} - \hat{\mathbf d}^{(p)} 
     \end{aligned}
     \end{equation}  is denoting the estimation error of the phase-center location. Specifically, the processing is done by modifying the measured  ${\bold A}^{(p)}$ as  
\begin{small}
     \begin{equation}
 \begin{aligned}
  & \tilde{\bold A}^{(p)}   \\
  & =  \exp\Biggl(  -j 2\pi \lambda^{-1}\begin{bmatrix} 
    \mathbf u^T_{0, 0} \hat{\mathbf d}^{(p)}  & \cdots &  \mathbf u^T_{0, \frac{(2N-1)\pi}{N}} \hat{\mathbf d}^{(p)} \\
     \vdots &  \vdots & \vdots\\
     \mathbf u^T_{\pi, 0} \hat{\mathbf d}^{(p)} &  \cdots &  \mathbf u^T_{\pi, \frac{(2N-1)\pi}{N}} \hat{\mathbf d}^{(p)} \\
    \end{bmatrix} \Biggr) \odot  {\bold A}^{(p)}.
    \end{aligned}
   \label{eq:Apreprocessing}
    \end{equation}
  \end{small} 
  The ${\bold A}^{(p)}$s in \eqref{eq:extending_3d_pattern} and \eqref{eq:aprime} are then replaced by the processed $\tilde{\bold A}^{(p)}$s to obtain the $\tilde{\bold Q}_{\text{H}}^{(p)}$s and $\tilde{\bold Q}_{\text{V}}^{(p)}$s using \eqref{eq:eadf_calculation}. 
    The purpose of \eqref{eq:Apreprocessing} is to compensate the fast phase rotations in $\mathbf A^{(p)}$ due to the misalignment $\mathbf d^{(p)}$, which is analogous to moving the signals in radio frequency to baseband. It can be known from \eqref{eq:compensation} that the maximum admissible angle steps for $\tilde{\bold A}^{(p)}$ is jointly determined by $\tilde{\mathbf d}^{(p)}$  and the intrinsic characteristic of $a_\text{0}^{(p)}$. With $||\tilde{\mathbf d}^{(p)}||$, or $\lambda^{-1} ||\tilde{\mathbf d}^{(p)}||$, being small, we can usually use much large angle steps for $\tilde{\bold A}^{(p)}$ compared to that of ${\bold A}^{(p)}$. Finally, the ``baseband'' signals need to be moved back to the high-frequency band. That is, with new $\tilde{\bold Q}_{\text{H}}^{(p)}$s and $\tilde{\bold Q}_{\text{V}}^{(p)}$s, \eqref{eq:poloreadf} can be modified as 
    \begin{small}
    \begin{equation}
      \begin{aligned}
          & \bold G (\theta, \phi) \\
          & = \Biggl(\begin{bmatrix} 
          \text{vec}^{T}(\tilde{\bold Q}_{\text{H}}^{(1)})  &  \text{vec}^{T}(\tilde{\bold Q}_{\text{V}}^{(1)}) \\
           \vdots &  \vdots \\
           \text{vec}^{T}(\tilde{\bold Q}_{\text{H}}^{(P)}) &  \text{vec}^{T}(\tilde{\bold Q}_{\text{V}}^{(P)})  \\
          \end{bmatrix} \begin{bmatrix} 
            \boldsymbol{\omega}_{\theta\phi}   & \mathbf 0 \\
            \mathbf 0 &  \boldsymbol{\omega}_{\theta\phi}   \\
            \end{bmatrix}\Biggr) \odot
            \begin{bmatrix} 
              \boldsymbol{\beta}_{\theta,\phi} & \boldsymbol{\beta}_{\theta,\phi} 
              \end{bmatrix}
       \end{aligned}
       \label{eq:poloreadf_modified}
       \end{equation}
      \end{small}
    for the polarimetric characterization of the array, where 
    \begin{equation}
      \begin{aligned}
        \boldsymbol{\beta}_{\theta,\phi} = \exp(j2\pi \lambda^{-1} [\mathbf u^T_{\theta, \phi} \hat{\mathbf d}^{(1)},\cdots, \mathbf u^T_{\theta, \phi} \hat{\mathbf d}^{(P)}]^{T}).
       \end{aligned}
       \end{equation}

\subsection{Obtaining the phase centers of the array elements}

By assuming that the phase center of each array element is located at its geometric center, it is possible to obtain $\hat{\mathbf d}^{(p)}$s by checking the array configuration and how the array is placed in the Cartesian system as shown in Fig.\,\ref{fig:array_measurement}. However, there are several drawbacks or limitations. \textit{i)} It may not be straightforward to find the geometric center of an array element. It is also possible that the phase center is not aligned with the geometric center.  \textit{ii)} The antenna array may not be visible. In this case, it is not easy to know the array geometry. \textit{iii)} It is generally hard to accurately measure the location and orientation of the array in the anechoic chamber. All these can lead to large errors in $\hat{\mathbf d}^{(p)}$s, i..e., large $||\tilde{\mathbf d}^{(p)}||$s. Therefore, we propose a new method to estimate the phase centers. A prerequisite of the proposed method is that the responses of the antenna elements across a certain frequency band are measured in the anechoic chamber. Usually, wideband characterization of an antenna array is necessary, especially for mmWave and sub-THz communications with a large amount of available bandwidth \cite{mmWaveSounder,9115069}, which inherently meets the prerequisite. Even if only one frequency point is concerned for the characterization of the antenna array, it is worth measuring the responses at additional frequency points since the most time-consuming operation during the measurement is mechanically rotating the probe and/or the array. 

Let us denote the measured responses of the $p$-th antenna element at different frequency points and directions as $a^{(p)}(f,\theta, \phi)$. It can be expressed as
\begin{equation}
  \begin{aligned}
    a^{(p)}(f,\theta, \phi) = \alpha^{(p)}(f, \theta, \phi)\exp(-j2\pi f \tau^{(p)}_{\theta, \phi}) + n(f, \theta, \phi)
   \end{aligned}
   \label{eq:26}
   \end{equation} where 
   \begin{equation}
    \begin{aligned}
      {\tau}^{(p)}_{\theta, \phi} =  \Delta \tau^{(p)} - c^{-1}\mathbf u^T_{\theta, \phi} \mathbf d^{(p)} 
     \end{aligned}
     \end{equation} 
     denotes the propagation delay with $c$ being the speed of light and $\Delta \tau^{(p)}$ being a fixed delay offset caused by the additional propagation in the connecting cable and circuit, $\alpha^{(p)}(f, \theta, \phi)$ is the residual complex amplitude, and $n$ is additive white Gaussian noise. Given the measured responses of $S$ frequency points, i.e., $[f_1,\cdots, f_S]$, at the direction $(\theta, \phi)$, the propagation delay at this direction can be estimated as
     \begin{equation}
    \begin{aligned}
      \hat{\tau}^{(p)}_{\theta, \phi} = \max_{\tau}  \enspace \biggl|  \sum_{s=1}^{S} a^{(p)}(f_s,\theta, \phi)   \exp(j2\pi f_s \tau)  \biggr|^2. 
     \end{aligned}
     \label{eq:mle_tau}
     \end{equation} 
     It can be seen that \eqref{eq:mle_tau} is a maximum-likelihood estimate if $\alpha^{(p)}(f_s, \theta, \phi)$ is independent on $f_s$, which is almost true if the considered frequency band is narrow. In the wideband case, although the amplitude variation of $\alpha^{(p)}(f_s, \theta, \phi)$ with respect to $f_s$ could be large, its phase variation should be generally small in the main-coverage directions, where \eqref{eq:mle_tau} can still be used for the estimation of ${\tau}^{(p)}_{\theta, \phi}$. It is worth noting that one should avoid estimating ${\tau}^{(p)}_{\theta, \phi}$ in the directions with much weaker antenna gains where the phase of $\alpha^{(p)}(f_s, \theta, \phi)$ could change irregularly leading to systematic estimation errors. With the propagation distances estimated at different directions, the phase center of the $p$-th antenna can then be estimated using the least-square principle as 
     \begin{equation}
      \begin{aligned}
        (\hat{\mathbf d}^{(p)}, \Delta {\hat\tau}^{(p)} ) = \max_{(\mathbf d, \Delta\tau)} \enspace  \sum_{\theta, \phi} \biggl| c^{-1}\mathbf u^T_{\theta, \phi}{\mathbf d} -   \Delta\tau  - \hat{\tau}^{(p)}_{\theta, \phi}   \biggr|^2. 
       \end{aligned}
       \label{eq:mle_d}
       \end{equation} After obtaining $\hat{\mathbf d}^{(p)}$s for all the $P$ elements, the enhanced EADF can be applied to the large-scale array. 


\section{Measurement verification\label{sec:measurement_verification}}

In this section, a measurement campaign using a mmWave antenna array at the frequency band from 27\,GHz to 30\,GHz is introduced, and the application of the proposed principle of enhanced EADF is verified with the measurement data. 

\subsection{Measurement campaign}

\begin{figure}
  \centering
  \psfrag{a}[c][c][0.55]{\color{white} Measured elements}
  \psfrag{control}[r][r][0.55]{\colorbox{white!30}{Switch-control cables}}
  \includegraphics[width=0.45\textwidth]{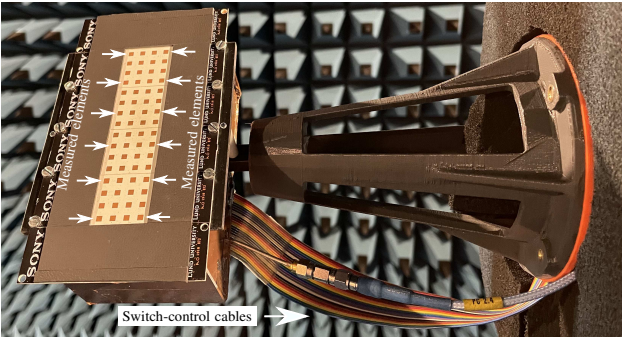}
  \caption{The planar array applied in the measurement. \label{fig:planar_array}}
 \end{figure}

\begin{figure}
  \centering
  \psfrag{x}[c][c][0.8]{\color{white} X}
  \psfrag{y}[c][c][0.8]{\color{white} Y}
  \psfrag{z}[c][c][0.8]{\color{white} Z}
  \psfrag{S}[l][l][0.55]{\colorbox{white!30}{Generate switch-control signals}}
  \psfrag{Probe}[l][l][0.55]{\colorbox{white!30}{Probe antenna}}
  \includegraphics[width=0.45\textwidth]{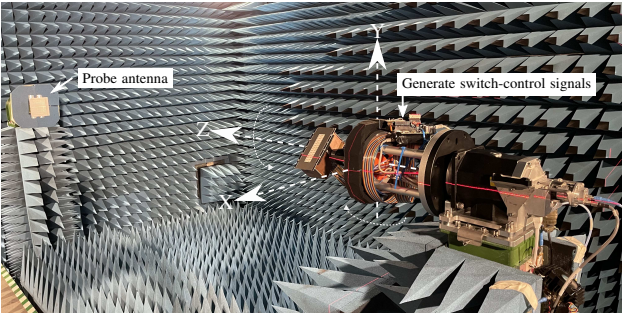}{}
  \caption{The measurement setup in the anechoic chamber. \label{fig:planar_array_in_ac}}
 \end{figure}

As illustrated in Fig.\,\ref{fig:planar_array}, the antenna array applied in the measurement is planar. There are 64 metallic squares on the panel each corresponding to two co-located H- and V-polarized antenna elements. That is, the planar array consists of 128 elements in total. Moreover, the spacing between neighbouring squares is around 6\,mm. This array is a switched array that was designed as a transmitter array of a mmWave channel sounder to measure dynamic mmWave massive MIMO radio channels \cite{mmWaveSounder}. A single element is activated by a 7-bits digital control signal from the rainbow cables as illustrated in Fig.\,\ref{fig:planar_array}. For more details of the array, readers are referred to \cite{mmWaveSounder}. 

Fig.\,\ref{fig:planar_array_in_ac} illustrates the measurement setup in the anechoic chamber, with the coordinate system also plotted. A high-gain antenna \cite{8958082} was fixed at its position as a probe. The planar array was installed on a piece of machinery with rotation functions, and the center of the array panel was around 20\,cm (38$\lambda$ at 28.5\,GHz) far from the origin along the $z$-axis. The distance from the probe to the origin of the coordinate system was 3\,m, which is larger than the Fraunhofer distance of the array. The machinery can rotate around the $y$-axis, i.e., rotate the coordinate system around the $y$-axis, to change the zenith angle of the probe, and can rotate around the $z$-axis to change the azimuth angle of the probe. Both the zenith step and the azimuth step were set as 1.5$^\circ$, i.e., $M=120$ and $N=120$, respectively. At each direction, twelve V-polarized antenna elements, as indicated by the white arrows in Fig.\,\ref{fig:planar_array}, were switched and measured. For each element, frequency responses at $S=301$ points that span from 27\,GHz to 30\,GHz were recorded, to cover the working frequency range of the array. 
It took around 19 hours to finish the measurement for a whole sphere, and a major part of the time was for mechanical rotations. In the sequel, we investigate the EADFs at the center frequency point 28.5\,GHz.

 \subsection{Estimation of phase centers}
  
 \begin{figure}
  \centering
  \psfrag{zenith}[c][c][0.6]{$\theta$ $[^\circ]$}
  \psfrag{azimuth}[c][c][0.6]{$\phi$ $[^\circ]$ }
  \psfrag{Gain [dB]}[c][c][0.6]{Gain [dB]}
  \subfigure[]{\includegraphics[width=0.49\textwidth]{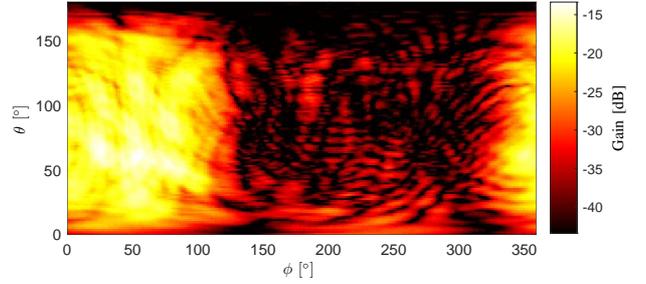}}
  \psfrag{zenith}[c][c][0.6]{$\theta$ $[^\circ]$}
  \psfrag{azimuth}[c][c][0.6]{$\phi$ $[^\circ]$ }
  \psfrag{Phase}[c][c][0.6]{Phase [$^\circ$]}
  \addnew{{
  \subfigure[]{\includegraphics[width=0.49\textwidth]{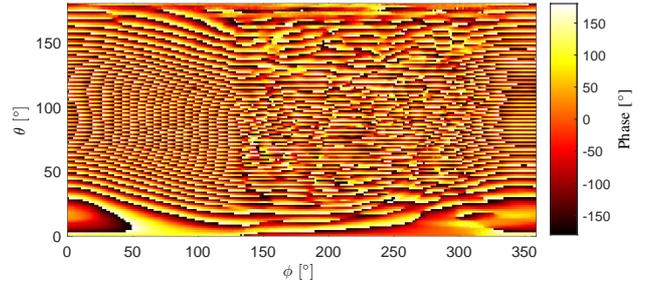}}
  }
  }
  \caption{The originally measured radiation pattern $\mathbf A^{(1)}$ of the first antenna element of the array as illustrated in Fig.\,\ref{fig:planar_array}. \addnew{(a) Gain pattern. (b) Phase pattern.}\label{fig:measured_pattern1}}
 \end{figure}

\begin{figure}
  \centering
  \psfrag{zenith}[c][c][0.6]{$\theta$ $[^\circ]$}
  \psfrag{azimuth}[c][c][0.6]{$\phi$ $[^\circ]$ }
  \psfrag{Estimated delay in [m]}[c][c][0.6]{Estimated delay [m]}
  \subfigure[]{\includegraphics[width=0.49\textwidth]{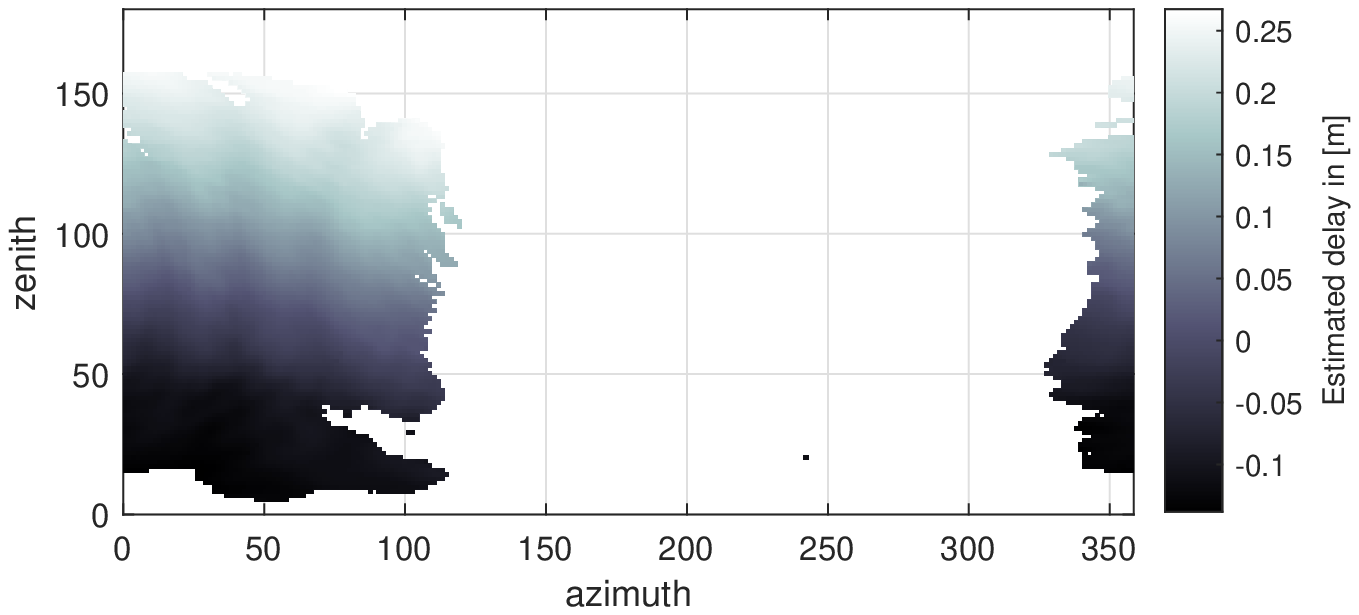}\label{fig:delay_results}}
  \psfrag{zenith}[c][c][0.6]{$\theta$ $[^\circ]$}
  \psfrag{azimuth}[c][c][0.6]{$\phi$ $[^\circ]$ }
  \psfrag{Fitted delay in [m]}[c][c][0.6]{Fitted delay [m]}
  \subfigure[]{\includegraphics[width=0.49\textwidth]{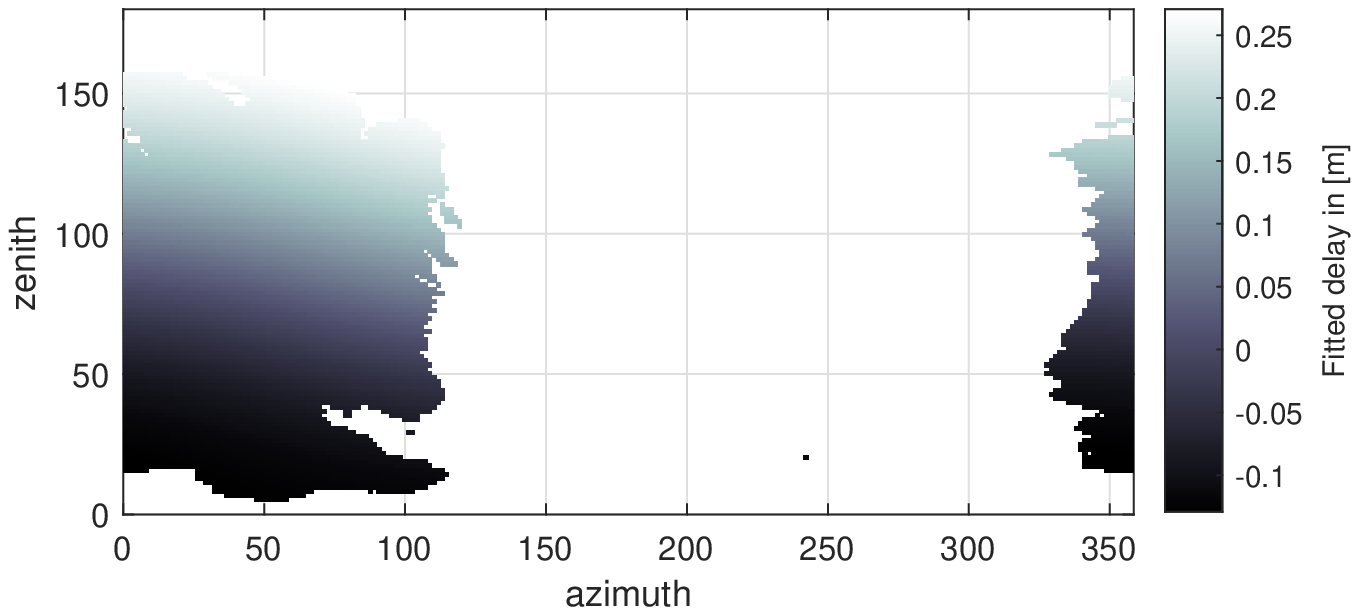}\label{fig:fitted_delays}}
  \caption{Propagation delays. (a) The estimated propagation delays using measured $\mathbf A^{(1)}$ according to \eqref{eq:mle_tau}. (b) The fitted delays using $\hat{\mathbf d}^{(1)}$ and $\Delta \hat{\tau}^{(1)}$ obtained from \eqref{eq:mle_d}. \label{fig:delays}}
 \end{figure}

\begin{figure}
  \centering
  \psfrag{xdistance [m]}[c][c][0.8]{$x$ [m]}
  \psfrag{ydistance [m]}[r][r][0.8]{$y$ [m]}
  \psfrag{zdistance [m]}[c][c][0.8]{$z$ [m]}
  \includegraphics[width=0.49\textwidth]{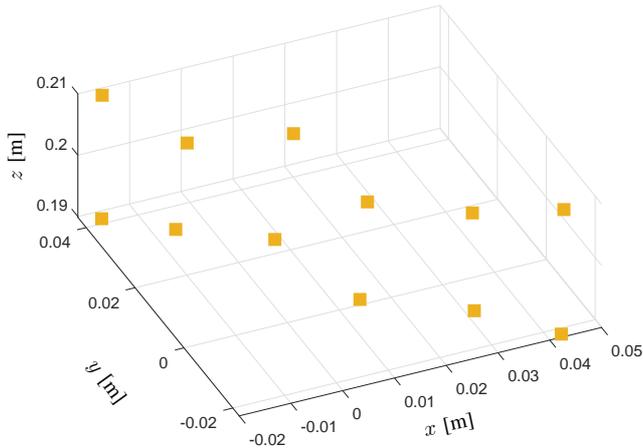}
    \caption{The estimated phase centers of the measured 12 antenna elements of the array. \label{fig:estimated_phase_centers}}
 \end{figure}

Fig.\,\ref{fig:measured_pattern1} illustrates, as an example, the measured pattern of the first antenna element at 28.5\,GHz, i.e., $\mathbf A^{(1)}$. \addnew{It can be observed from Fig.\,\ref{fig:measured_pattern1}(b) that the phases at different directions change fast due to the phase center offset of this antenna element.} 
Fig.\,\ref{fig:delay_results} illustrates the estimated propagation delays  $\hat{\tau}^{(1)}_{\theta, \phi}$ at different directions of the first antenna element according to \eqref{eq:mle_tau}. Note that only the main coverage, i.e., directions with gains no smaller than 13\,dB below the maximum gain, is considered for delay estimation, to avoid possible systematic errors that could happen in the weak-gain area. 
It can be observed that the propagation distance is changing from negative to positive values while the zenith angle is varying from 0 to $\pi$, which is consistent with the measurement setup as shown in Fig.\,\ref{fig:planar_array_in_ac}. The phase center is then estimated as $\hat{\mathbf d}^{(1)}$ according to the least-square principle \eqref{eq:mle_d}. Fig.\,\ref{fig:fitted_delays} illustrates the fitted delays using the estimated $\hat{\mathbf d}^{(1)}$ and $\Delta \hat{\tau}^{(1)}$. It can be observed that the least-square fitting is quite feasible. By performing the procedure for the 12 antenna elements respectively, their phase centers are obtained and depicted in Fig.\,\ref{fig:estimated_phase_centers}. It can be observed that the estimated phase centers are quite close to (although not exactly aligned with) the geometry centers of the 12 metallic patches on the panel as illustrated in Fig.\,\ref{fig:planar_array}. 

 \subsection{Performance verification of the enhanced EADF}

\begin{figure}
  \centering
  \psfrag{zenith}[c][c][0.6]{$\theta$ $[^\circ]$}
  \psfrag{azimuth}[c][c][0.6]{$\phi$ $[^\circ]$ }
  \psfrag{Phase}[c][c][0.6]{Phase [$^\circ$]}
  \addnew{{
  \includegraphics[width=0.47\textwidth]{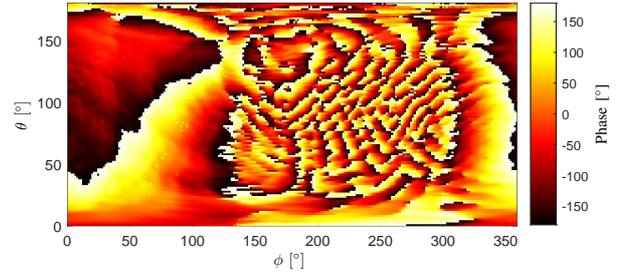}
  }
  }
  \caption{\addnew{The phase pattern of the first antenna element after phase compensation according to the estimated phase center, i.e., the phase pattern of $\tilde{\bold A}^{(1)}$. \label{fig:aftercompensation}}}
 \end{figure}

\begin{figure}
  \centering
  \psfrag{zenith frequency}[c][c][0.6]{Zenithal spatial frequency}
  \psfrag{azimuth frequency}[c][c][0.6]{Azimuthal spatial frequency}
  \psfrag{Power [dB]}[c][c][0.6]{Normalized power [dB]}
  \subfigure[]{\includegraphics[width=0.24\textwidth]{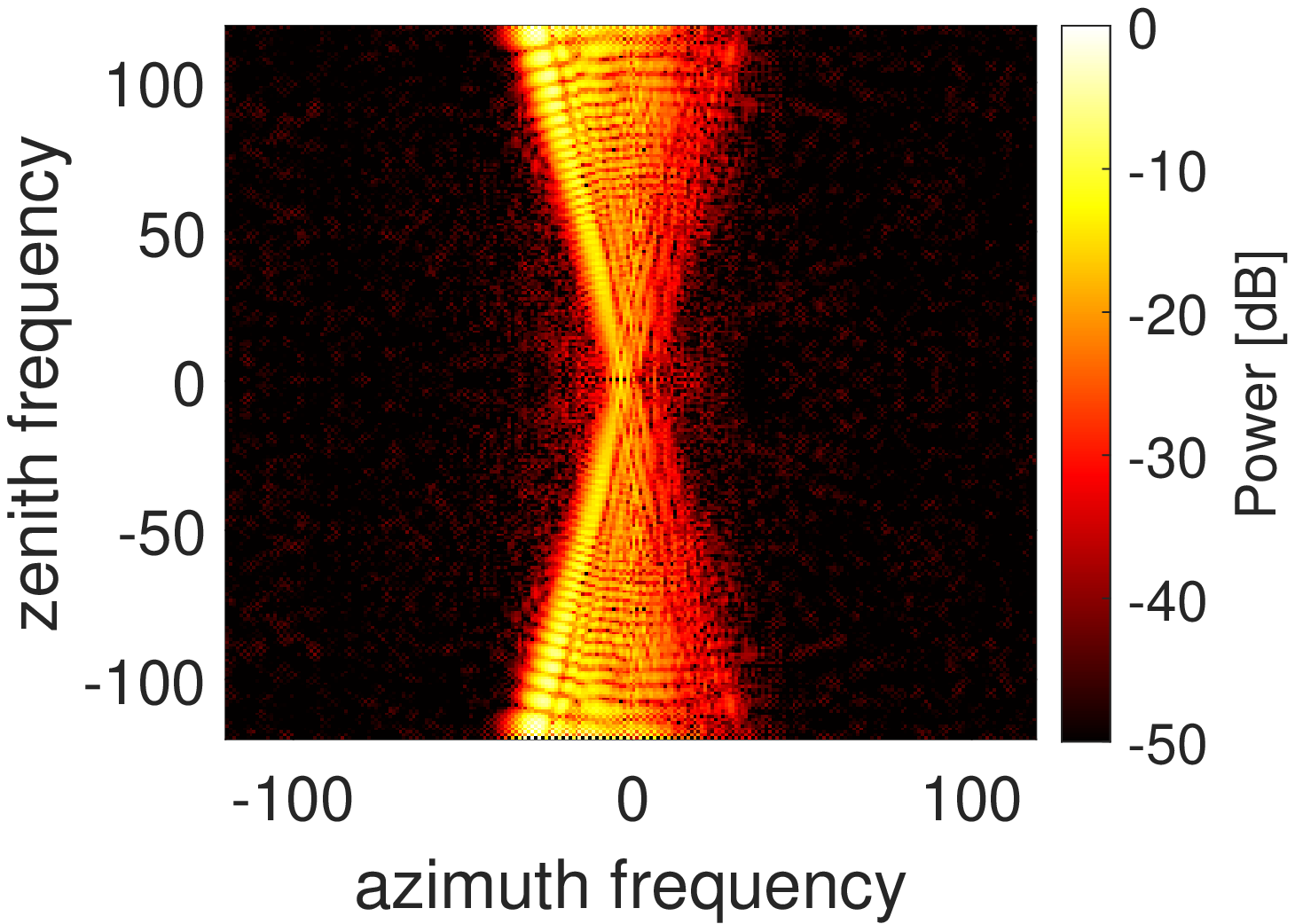}\label{fig:previousEADF}}
  \psfrag{zenith frequency}[c][c][0.6]{Zenithal spatial frequency}
  \psfrag{azimuth frequency}[c][c][0.6]{Azimuthal spatial frequency}
  \psfrag{Power [dB]}[c][c][0.6]{Normalized power [dB]}
  \subfigure[]{\includegraphics[width=0.24\textwidth]{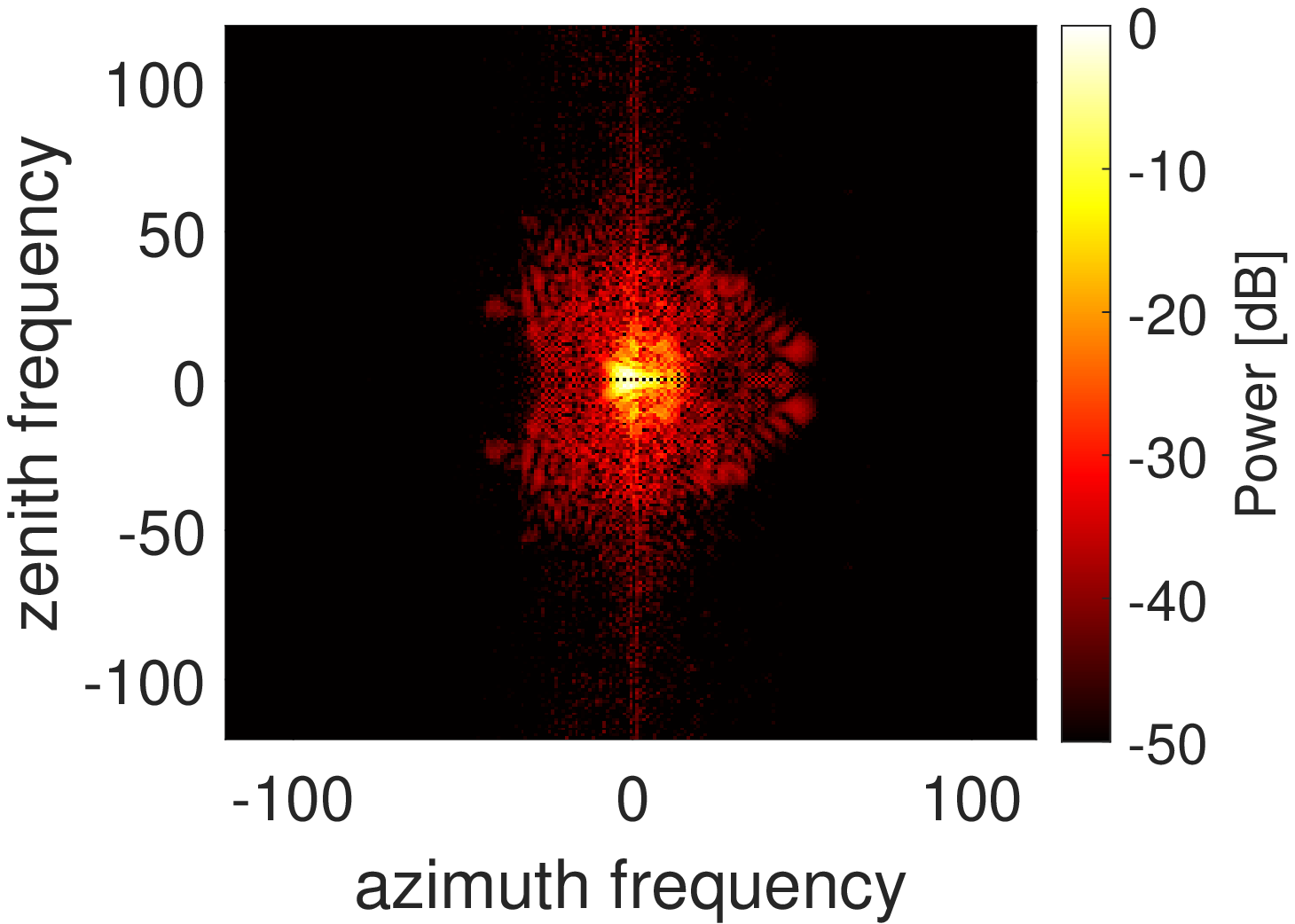}\label{fig:newEADF}}
  \caption{Normalized EADFs for the first antenna. (a) The EADF $\mathbf Q^{(1)}$ obtained using the original measured $\mathbf A^{(1)}$. (b) The enhanced EADF $\tilde{\mathbf Q}^{(1)}$ obtained using $\tilde{\bold A}^{(1)}$. } 
 \end{figure}

 As a comparison between the conventional EADF and the proposed enhanced EADF, \addnew{Fig.\,\ref{fig:aftercompensation} illustrates the phase pattern of the first antenna element after phase compensation according to the estimated phase center, i.e., the phase pattern of $\tilde{\bold A}^{(1)}$. It can be observed that the fast phase rotation of the originally measured $\mathbf A^{(1)}$ in the main coverage area as illustrated in Fig.\,\ref{fig:measured_pattern1}(b) is well compensated.} 
 Moreover, Figs.\,\ref{fig:previousEADF} and \ref{fig:newEADF} illustrate the normalized EADF power spectra of the originally measured $\mathbf A^{(1)}$ and the processed $\tilde{\bold A}^{(1)}$ by \eqref{eq:Apreprocessing} with the estimated $\hat{\mathbf d}^{(1)}$, respectively. It can be clearly observed from Fig.\,\ref{fig:previousEADF} that there are strong components spanning the whole zenith-spatial domain, which is an indication that the angle step of 1.5$^\circ$ is not adequate to sample the original pattern $\mathbf A^{(1)}$ that has high local spatial frequencies due to the distance offset of the phase center. This is consistent with the analysis of \eqref{eq:max_zenith_step} that the zenith step should be (further) smaller than $\frac{1}{2\times38}$\,radians, i.e., 0.75$^\circ$, to adequately sample the high-spatial frequency components in conventional EADF. In contrast, by compensating the fast phase-rotation in $\mathbf A^{(1)}$, the energy of the new EADF as illustrated in Fig.\,\ref{fig:newEADF} is concentrated around the low spatial-frequency domain. This means that we can use much larger angle steps to sample the $\tilde{\mathbf A}^{(1)}$ compared to that of $\mathbf A^{(1)}$.

 To evaluate the performance of EADF, we define a metric, i.e., relative error magnitude (REM), that is formulated as
 \begin{equation}
  \begin{aligned}
    \epsilon^{(p)}(\theta, \phi) = \Bigl|\frac{ {a}^{(p)}(\theta, \phi) - \hat{a}^{(p)}(\theta, \phi)  }{{a}^{(p)}(\theta, \phi)} \Bigr|,
   \end{aligned}
   \label{eq:error_magnitude}
   \end{equation} 
  where $\hat{a}^{(p)}(\theta, \phi)$ is the recovered response using the EADF, and ${a}^{(p)}(\theta, \phi)$ is the ground-truth response. In the measurement campaign, we succeeded in achieving a signal-to-noise ratio (SNR) of around 60\,dB at 28.5\,GHz in the boresight direction, thus the measured responses are approximated as the ground-truth responses. Moreover, since the raw $\mathbf A$s were measured with angle steps of 1.5$^\circ$, we can re-sample them to construct new $\mathbf A$s with larger angle steps that are chosen from 
   $[3, 4.5, 6, 7.5, 9, 12, 15, 18, 22.5, 30, 36, 45, 60]$\,\,degrees. With new $\mathbf A$s, the responses at the other measured directions can be recovered using EADF, and the REMs can be calculated.

\begin{figure}
    \centering
    \addnew{{
    \psfrag{Gain [dB]}[c][c][0.6]{Gain [dB]}
    \psfrag{Elevation}[c][c][0.6]{$\theta$ $[^\circ]$ }
    \psfrag{measured}[l][l][0.6]{Measured}
    \psfrag{eadf}[l][l][0.6]{EADF}
    \psfrag{enhanced}[l][l][0.6]{Enhanced EADF}
    \includegraphics[width=0.44\textwidth]{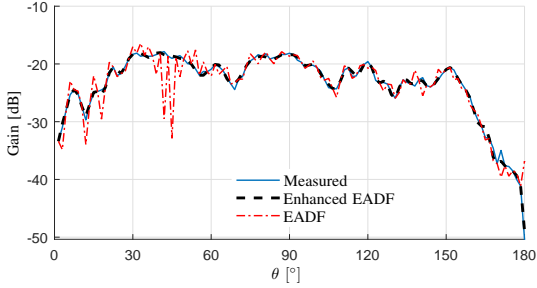}
    }}
    \caption{\addnew{The measured and reconstructed patterns at the azimuth cut of 7.5$^\circ$ for the first antenna element of the array.} \label{fig:patterncuts}}
   \end{figure}

\begin{figure}
    \centering
    \psfrag{zenith}[c][c][0.6]{$\theta$ $[^\circ]$}
    \psfrag{azimuth}[c][c][0.6]{$\phi$ $[^\circ]$ }
    \psfrag{Error [dB]}[c][c][0.6]{Relative error magnitude [dB]}
    \subfigure[]{\includegraphics[width=0.49\textwidth]{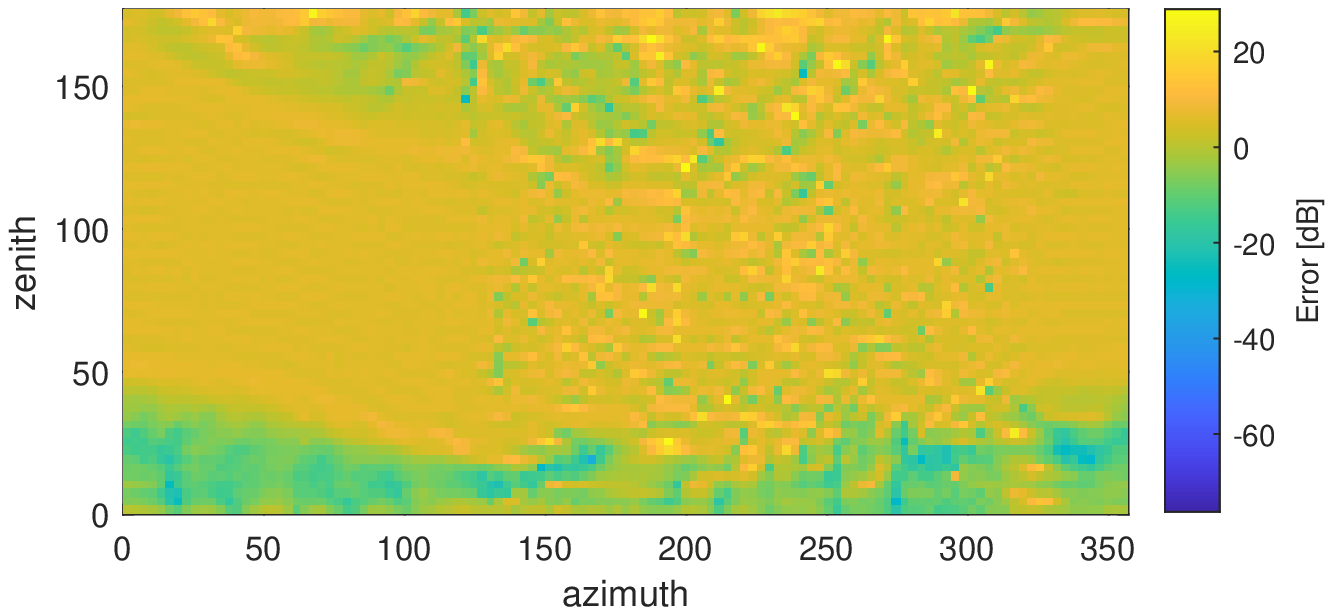}\label{fig:old_eadf_error}}
    \psfrag{zenith}[c][c][0.6]{$\theta$ $[^\circ]$}
    \psfrag{azimuth}[c][c][0.6]{$\phi$ $[^\circ]$ }
    \psfrag{Error [dB]}[c][c][0.6]{Relative error magnitude [dB]}
    \subfigure[]{\includegraphics[width=0.49\textwidth]{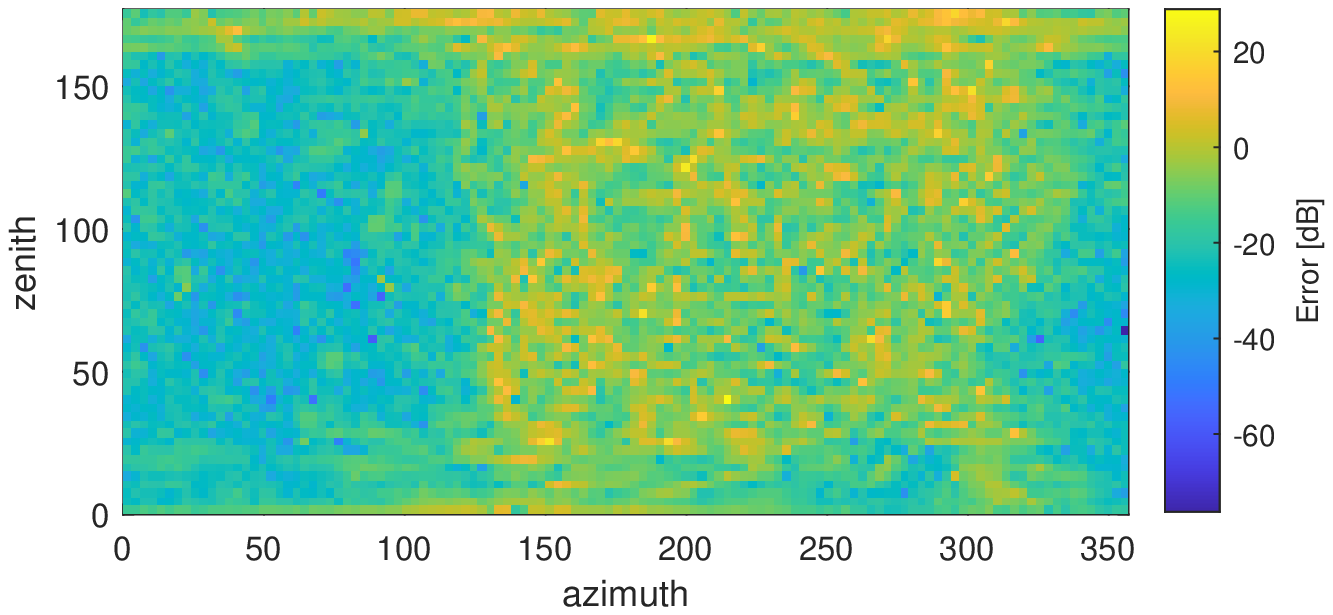}\label{fig:new_eadf_error}}
    \caption{The REM of the characterization using conventional EADF and the enhanced EADF for the first antenna element of the array, where $\mathbf A^{(1)}$ is re-sampled with 3$^\circ$ steps. (a) The conventional EADF. (b) The enhanced EADF.}
   \end{figure}

   \addnew{For example, Fig.\,\ref{fig:patterncuts} illustrates the measured pattern, reconstructed pattern from the conventional EADF, and reconstructed pattern from the enhanced EADF at the azimuth cut of 7.5$^\circ$ for the first antenna element, where a re-sampled $\mathbf A^{(1)}$ with angle steps of 3$^\circ$ is exploited in both the conventional EADF and the enhanced EADF. It can be observed that the reconstructed pattern using the enhanced EADF is quite consistent with the measurement, while the reconstructed pattern from the conventional EADF has significant errors due to the aliasing issue.} 
   Moreover, Fig.\,\ref{fig:old_eadf_error} illustrates the 3D REM for the first antenna element, where a re-sampled $\mathbf A^{(1)}$ with angle steps of 3$^\circ$ is exploited to recover the antenna responses using the conventional EADF. As expected, the REM is quite large as more than 10\,dB in most cases, because the sampling intervals are much higher than the admissible values. In contrast, by using the enhanced EADF, the REM, as illustrated in Fig.\,\ref{fig:new_eadf_error}, can be decreased significantly to be around -25\,dB in the main coverage of the element. This verifies the advantage of the enhanced EADF. In addition, it can be observed from Fig.\,\ref{fig:new_eadf_error} that in the back area of the antenna coverage, the REM can still be higher than 10\,dB. We postulate that it is mainly due to the fast amplitude variation that can be seen in Fig.\,\ref{fig:measured_pattern1} and the irregular phase rotation therein, which can result in high local spatial frequencies. Nevertheless, it is a minor issue since the main-coverage area is the focus of the characterization.

   \begin{figure}
    \centering
    \psfrag{data1}[l][l][0.6]{Enhanced EADF, $3^\circ$}
    \psfrag{data2}[l][l][0.6]{Enhanced EADF, $4.5^\circ$}
    \psfrag{data3}[l][l][0.6]{Enhanced EADF, $6^\circ$}
    \psfrag{data4}[l][l][0.6]{Enhanced EADF, $7.5^\circ$}
    \psfrag{data5}[l][l][0.6]{Enhanced EADF, $9^\circ$}
    \psfrag{data6}[l][l][0.6]{Enhanced EADF, $12^\circ$}
    \psfrag{data7}[l][l][0.6]{Enhanced EADF, $15^\circ$}
    \psfrag{data8}[l][l][0.6]{Enhanced EADF, $18^\circ$}
    \psfrag{data9}[l][l][0.6]{Enhanced EADF, $22.5^\circ$}
    \psfrag{data10}[l][l][0.6]{Enhanced EADF, $30^\circ$}
    \psfrag{data11}[l][l][0.6]{Enhanced EADF, $36^\circ$}
    \psfrag{data12}[l][l][0.6]{Enhanced EADF, $45^\circ$}
    \psfrag{data13}[l][l][0.6]{Enhanced EADF, $60^\circ$}
    \psfrag{data14}[l][l][0.6]{EADF, $3^\circ$}
    \psfrag{F(x)}[c][c][0.8]{Probability\,$(E\leq abscissa)$}
    \psfrag{x}[c][c][0.8]{Relative error magnitude $\epsilon$ [dB]}
    \includegraphics[width=0.48\textwidth]{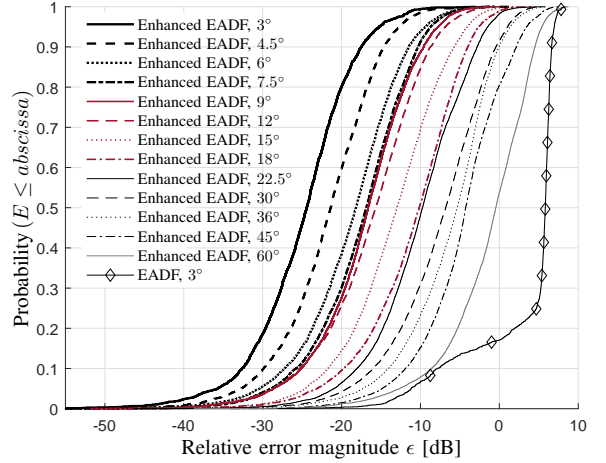}
      \caption{\color{black}{The CDFs of REMs at the main coverage area of the array in the measurement campaign as illustrated in Fig.\,\ref{fig:planar_array_in_ac}. The size of the array is around $3\lambda \times 15 \lambda$ at 28.5 GHz. The center of the array panel was around $38\lambda$ from the chamber origin along $z$-axis. The performance curve of the conventional EADF with angle steps of 3$^\circ$ is a baseline. The performance of enhanced-EADF with different angle steps from 3$^\circ$ to 60$^\circ$ are also shown.} \label{fig:rem_cdfs}}
   \end{figure}

   Fig.\,\ref{fig:rem_cdfs} illustrates the cumulative distribution functions (CDFs) of REMs of all the 12 antennas at their main-coverage area, the area as illustrated in Fig.\,\ref{fig:delay_results}, using the enhanced EADF and re-sampled $\mathbf A$s with different angle steps. 
   The CDF of REMs using the conventional EADF with re-sampled $\mathbf A$s of 3$^\circ$ angle steps is also plotted as a baseline. It can be observed that the enhanced EADF works well to recover the array responses with small errors. 
   Moreover, the performance of enhanced EADF is decreasing with increasing angle steps, which is reasonable. However, the performance of the enhanced EADF, even with sampling steps of 60$^\circ$, is better than that of the conventional EADF with angle steps of only $3^\circ$. This demonstrates the advantage of applying the enhanced EADF for the characterization of large-scale antenna arrays. \addnew{For the case of our measurement campaign, angle steps can be increased from 0.75$^\circ$ required by the conventional EADF to around 4.5$^\circ$, or even larger, by the enhanced EADF to achieve an acceptable characterization of the array. This can significantly save measurement time (more than 36$\times$ time-saving) since much fewer mechanical rotations are required in the anechoic chamber. Specifically, it takes around 36 hours with 0.75$^\circ$ angle steps for conventional EADF, while for the enhanced EADF it is only around 1 hour with 4.5$^\circ$ angle steps.} Moreover, the computational complexity of the EADF can also be decreased significantly because the sizes of measured $\mathbf A$s and the corresponding $\mathbf Q$s can be much smaller.


\section{Conclusions\label{sect:conclusions}}
In this paper, we presented the effective aperture distribution function (EADF) applied to characterizing antenna arrays. As the size of an array becomes large, we showed that the conventional EADF can lead to a fatal issue. That is, the 3D radiation patterns of the antenna elements of the array must be sampled with very small angle steps, resulting in an expensive measurement effort and computationally complex EADF for the array. To solve this problem, we proposed an enhanced EADF. By estimating the locations of the antenna elements' phase centers, the fast phase rotations that cause high local spatial frequencies can be compensated. In such a way, we can achieve spatial sampling intervals that are intrinsically determined by the pattern of each element and are generally much larger. The measurement results of a mmWave antenna array at 28.5\,GHz showed that the enhanced EADF can have much better performance even with large angle steps such as 60$^\circ$ compared to that of the conventional EADF with angle steps of only 3$^\circ$. This demonstrates the necessity and effectiveness of the proposed enhanced EADF, especially for large-scale antenna arrays that are common in 5G and beyond communications.

\section{Acknowledgements}

The authors would like to thank Martin Nilsson for his assistance in preparing the measurement setup.    




\setlength{\itemsep}{10em}
\renewcommand{\baselinestretch}{1}
\bibliographystyle{IEEEtran}
\bibliography{reference}

\begin{IEEEbiography}[{\includegraphics[width=1in,height=1.25in,clip,keepaspectratio]{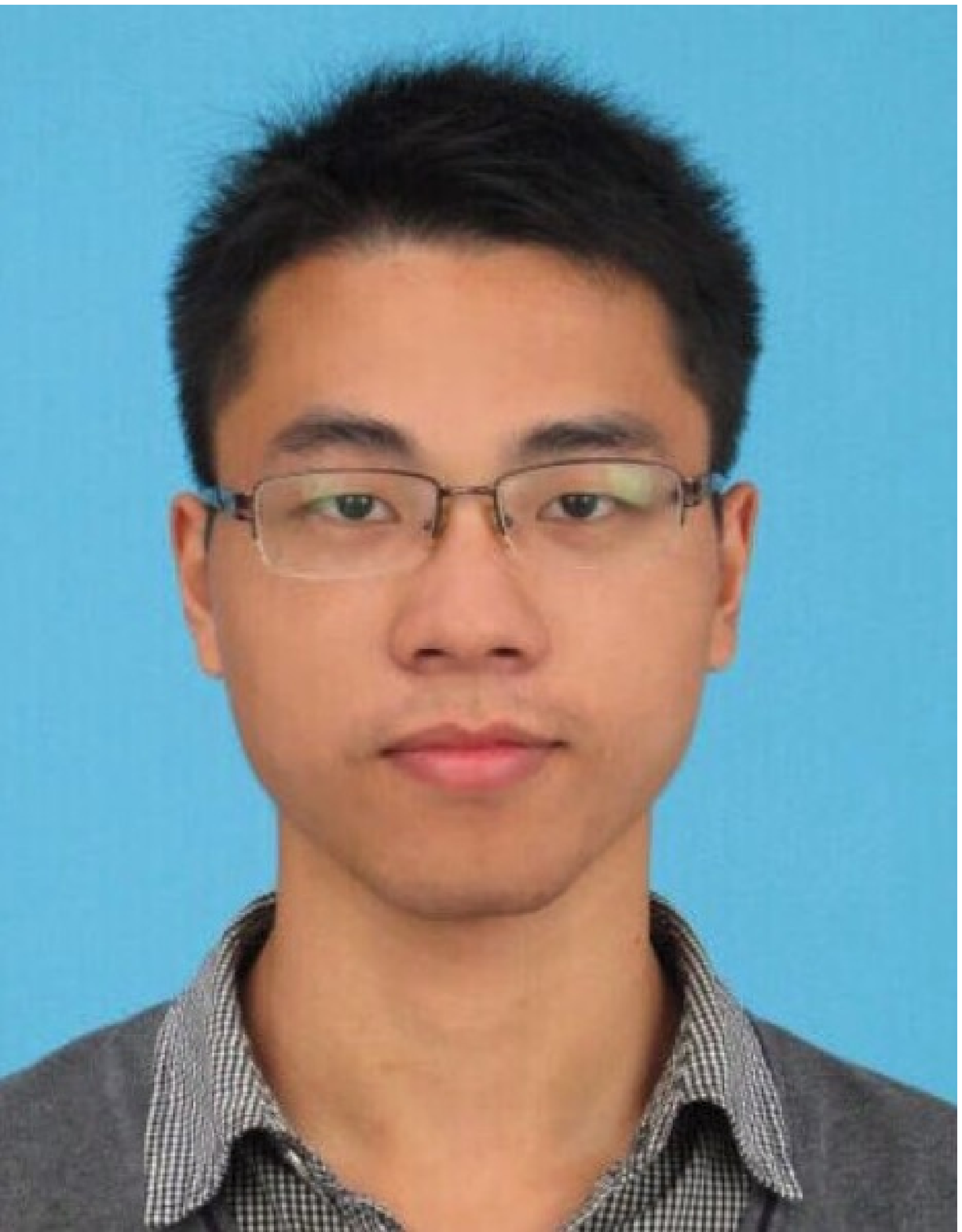}}]{Xuesong Cai} (Senior Member, IEEE) received the B.S. degree and the Ph.D.
  degree (with distinction) from Tongji University,
  Shanghai, China, in 2013 and 2018, respectively. In
  2015, he conducted a three-month internship with
  Huawei Technologies, Shanghai, China. He was also
  a Visiting Scholar with Universidad Politécnica de
  Madrid, Madrid, Spain in 2016. From 2018-2022,
  he conducted several postdoctoral stays at Aalborg
  University and Nokia Bell Labs, Denmark, and
  Lund University, Sweden. He is currently an Assistant
  Professor in Communications Engineering and a Marie Skłodowska-Curie Fellow at
  Lund University, closely cooperating with
  Ericsson and Sony. His research interests include radio propagation, high-resolution
  parameter estimation, over-the-air testing, resource optimization, and radio-based localization for 5G/B5G wireless systems. 

  Dr. Cai was a recipient of the China National Scholarship (the highest honor for Ph.D. Candidates) in 2016, the Outstanding Doctorate Graduate awarded by the Shanghai Municipal Education Commission in 2018, the Marie Skłodowska-Curie Actions (MSCA) ``Seal of Excellence'' in 2019, the EU MSCA Fellowship (ranking top 1.2\%, overall success rate 14\%) and the Starting Grant (success rate 12\%) funded by the Swedish Research Council in 2022. He was also selected by the ``ZTE Blue Sword-Future Leaders Plan'' in 2018 and the ``Huawei Genius Youth Program'' in 2021.
  
  
  
  \end{IEEEbiography}
  
  \begin{IEEEbiography}[{\includegraphics[width=1in,height=1.25in,clip,keepaspectratio]{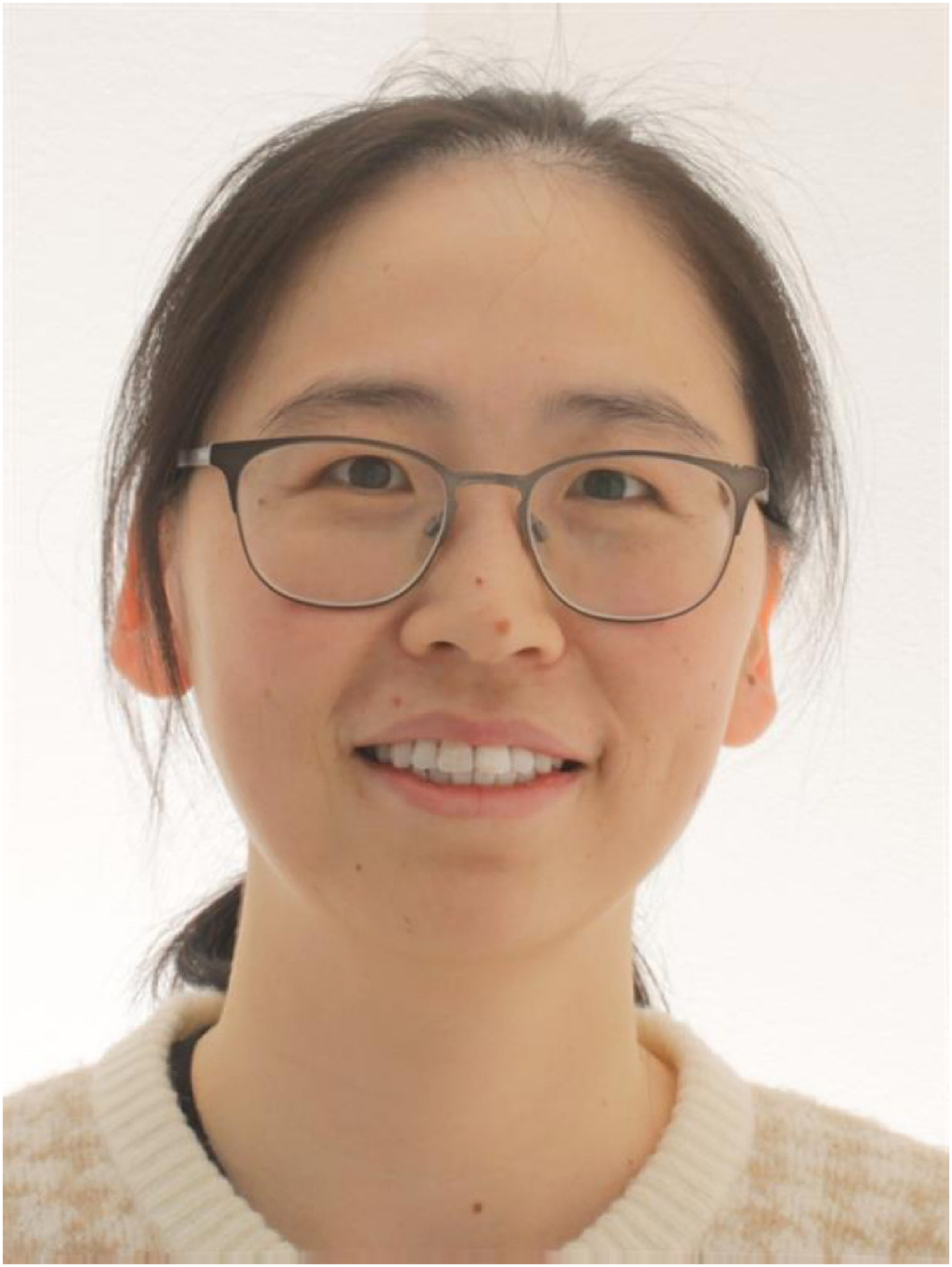}}]{Meifang Zhu} is a Senior Researcher at Ericsson AB, Lund, Sweden. Her interests include radio propagation, radio-based localization, 5G and beyond wireless communications Standardization and machine learning. 
  \end{IEEEbiography}

  \begin{IEEEbiography}[{\includegraphics[width=1in,height=1.25in,clip,keepaspectratio]{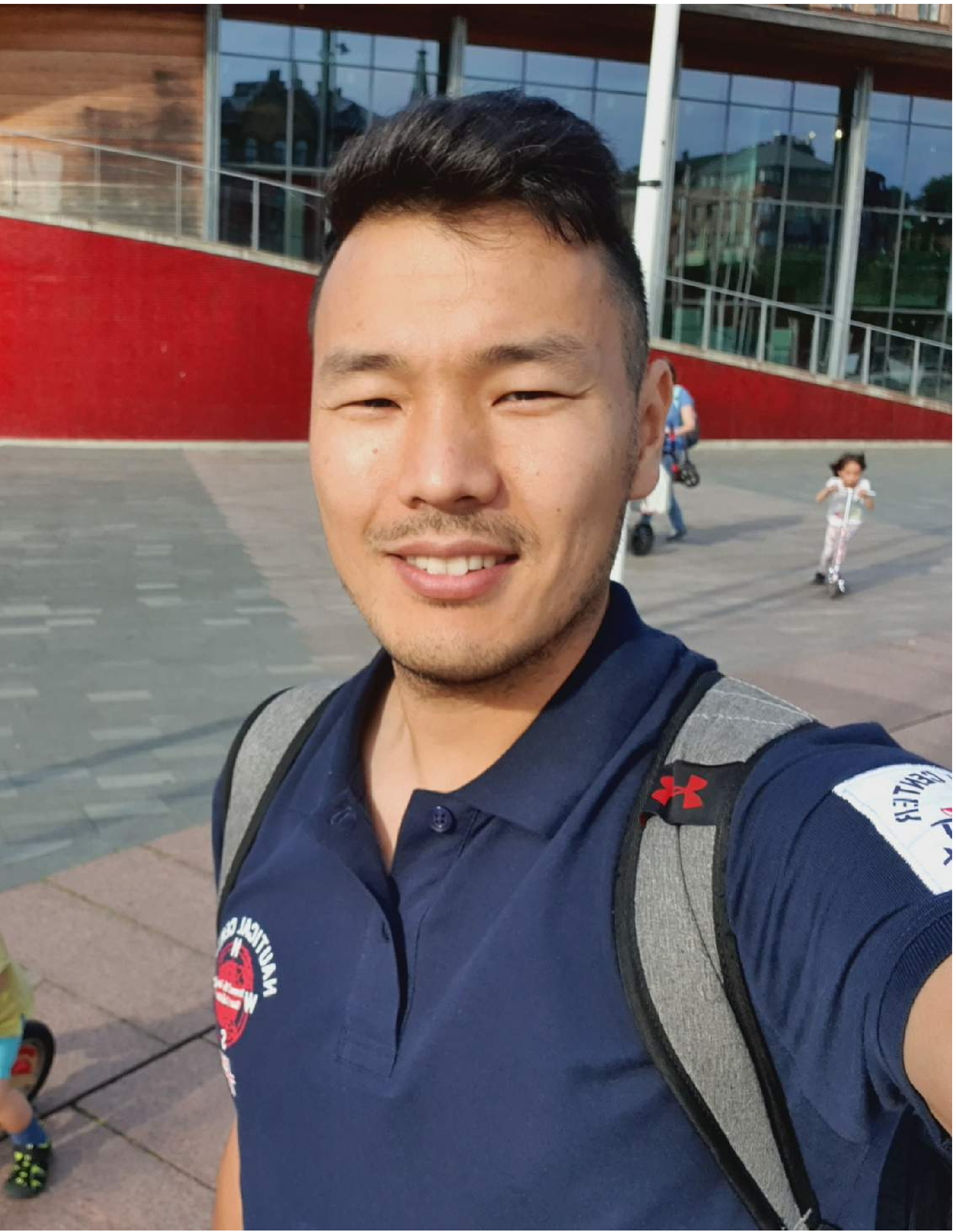}}]{Aleksei Fedorov} is a research engineer specializing in developing real-time simulators using the Unity game engine for wireless communication systems, including 5G, V2X, and LTE. He earned his PhD degree in wireless communications from Otago University in 2019 and holds a Master’s degree in theoretical mechanics from Lomonosov’s Moscow State University.
    He became a part of the EIT department at Lund University in 2019. Currently, his work is centered on creating interactive real-time 3D channel simulators for V2X, 5G, and emerging 6G applications.
Additionally, Aleksei Fedorov dedicates time to educating the next generation of professionals in the field by lecturing wireless communication-related courses for master's students at EIT. His work signifies a balanced mix of academic rigor, practical application, and a commitment to knowledge sharing in the field of wireless communication systems.

  \end{IEEEbiography}

  \begin{IEEEbiography}[{\includegraphics[width=1in,height=1.25in,clip,keepaspectratio]{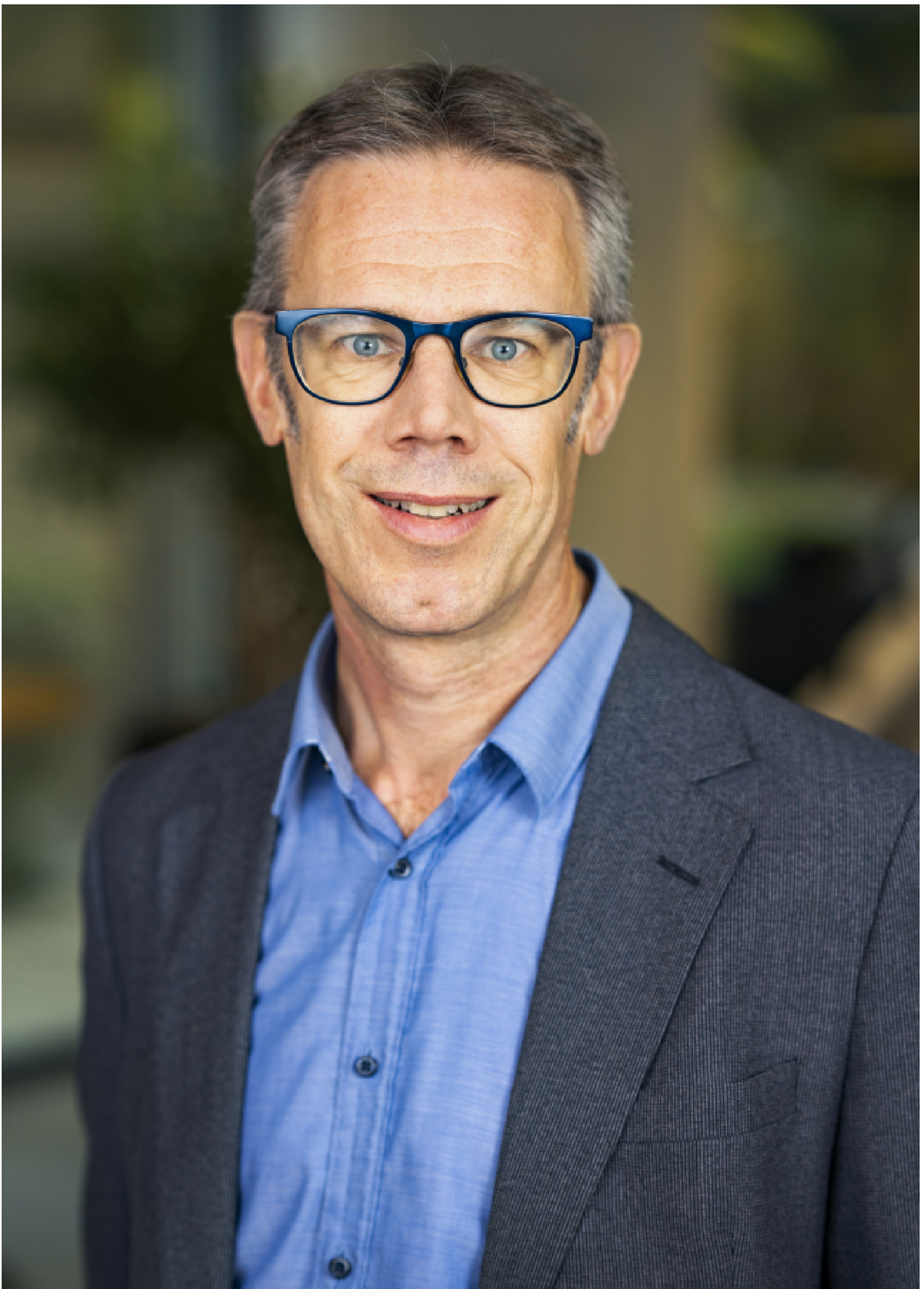}}]{Fredrik Tufvesson} (Fellow, IEEE) 
  received the Ph.D. degree from Lund University, Lund, Sweden, in 2000.
  
  After two years at a startup company, he joined the Department of Electrical and Information Technology, Lund University, where he is currently a Professor of radio systems. He has authored around 100 journal articles and 150 conference papers. His main research interest is the interplay between the radio channel and the rest of the communication system with various applications in 5G/B5G systems, such as massive multiple-input multiple-output (MIMO), mmWave communication, vehicular communication, and radio-based positioning.
  
  Dr. Tufvesson’s research has been awarded the Neal Shepherd Memorial Award for the Best Propagation Paper in the IEEE
  TRANSACTIONS ON VEHICULAR TECHNOLOGY and the IEEE Communications Society Best Tutorial Paper Award.

  \end{IEEEbiography}

\end{document}

%% file: newcommands_thesis2.tex
\newcommand{\azimuth}{\mbox{$\phi$}}

\newcommand{\rem}[1]{}

